\documentclass{aastex62}

\usepackage{lineno}

\usepackage{multirow}

\received{January 1, 2018}
\revised{January 7, 2018}
\accepted{\today}

\submitjournal{ApJ}

\shorttitle{The Tilt Angle of SG and the Properties of the Next SC}
\shortauthors{Gao \& Xu}

\begin{document}

\title{Relationship Between the Tilt Angle of Sunspot Group and the Properties of the Next Solar Cycle}

\correspondingauthor{P. X. Gao}
\email{gaopengxin@ynao.ac.cn}

\author{P. X. Gao}
\affil{Yunnan Observatories, Chinese Academy of Sciences, Kunming 650216, People's Republic of China}
\affil{Yunnan Key Laboratory of the Solar Physics and Space Science, Kunming 650216, People's Republic of China}

\author{J. C. Xu}
\affil{Yunnan Observatories, Chinese Academy of Sciences, Kunming 650216, People's Republic of China}
\affil{Yunnan Key Laboratory of the Solar Physics and Space Science, Kunming 650216, People's Republic of China}

\begin{abstract}
Based on the data from the Kodaikanal and Mount Wilson observatories, we investigate the relationships of the tilt angle of sunspot group (SG), including the mean tilt angle and the tilt-angle scatter, during the declining phase with the parameters of the next solar cycle (SC). The main findings are summarized in the following three points. (1) During the declining phase, the correlation between the mean tilt angle and the tilt-angle scatter is statistically insignificant. (2) Six quantities measured during the declining phase show significant anti-correlations with the strength and amplitude of the next SC, and positive correlations with the duration of the ascending phase of the next SC: the standard deviation of tilt angles, the root-mean-square tilt angle, the mean absolute value of tilt angles, the area-weighted absolute value of tilt angles, the latitude-weighted absolute value of tilt angles, and the area- and latitude- weighted absolute value of tilt angles. (3) The correlations of the mean tilt angle, the area-weighted tilt angle, the latitude-weighted tilt angle, and the area- and latitude- weighted tilt angle during the declining phase with the strength, amplitude, and duration of the ascending phase of the next SC are statistically insignificant. These findings demonstrate that the modulation of parameters of the next SC by the tilt-angle scatter during the declining phase plays a vital role in regulating SC variability.
\end{abstract}


\section{Introduction} \label{sec:intro}

The tilt angle of sunspot group (SG) is a crucial parameter in the conversion of the toroidal magnetic field to poloidal magnetic field, which is at the heart of the Babcock-Leighton-type dynamo model \citep{babcock61,leighton69}. It governs the conversion efficiency: in general, the larger the tilt angle, the greater the contribution to the poloidal field \citep{mazumder19,jiao21}. Thus, the tilt angle of SG is considered to be a direct observational determination of the principal driver in the solar dynamo \citep{tlatov13}.

The relationships of the tilt angle of SG with the parameters of the same solar cycle (SC) have been studied extensively in recent years based on the average tilt angle. For the first time, \cite{dasiespuig10} (and their corrigendum \cite{dasiespuig13}) found that $\langle\alpha\rangle/\langle|\lambda|\rangle$, where $\alpha$ and $|\lambda|$ were the tilt angle and the unsigned latitude, $\langle\alpha\rangle$ and $\langle|\lambda|\rangle$ were their SC-averaged values, had an anti-correlation with the amplitude of the same SC, with correlation coefficients of $r = -0.82$ at 98\% confidence level and $r = -0.80$ at 90\% confidence level for the Kodaikanal (KK) and Mount Wilson (MW) data, respectively. The SC-averaged tilt angle reported by \cite{dasiespuig10} shows significant anti-correlations with the strength and amplitude of the same SC at least for the KK data, while, for the MW data, the probabilities suggest that the anti-correlations are due to chance. It is difficult to reach a consensus on whether the tilt angle of SG shows an anti-correlation with the amplitude of the same SC \citep{ivanov12,mcclintock13}.

It has been widely accepted that the tilt angle of SG tends to increase with increasing latitude (Joy's law) \citep{hale19}. The linear equation,
\begin{equation}
\alpha=T_{\text{lin}}\times|\lambda|,
\end{equation}
is widely used to describe Joy's law. However, at higher latitudes, the increasing trend of tilt angle with increasing latitude tends to slow down gradually, or there may even be a decreasing trend with increasing latitude \citep{howard91,tlatov13,baranyi15,nagovitsyn21}. Thus, the square-root equation,
\begin{equation}
\alpha=T_{\text{sqr}}\times|\lambda|^{0.5},
\end{equation}
is also used to describe Joy's law \citep{cameron10,jiang20,jiao21}. \cite{jiang20} and \cite{jiao21} showed that the tilt coefficient, $T_{\text{lin}}$ or $T_{\text{sqr}}$, had an anti-correlation with the amplitude of the same SC. The tilt coefficient for linear form ($T_{\text{lin}}$) is same as the slope of Joy's law. There may be a nonlinear mechanism to modulate the generation of poloidal field \citep{lemerle17,karak17,jiang20,jiao21}.

The polar field at solar minimum has a strong correlation with the amplitude of the next SC \citep{schatten78,schatten05,svalgaard05,wang05,jiang07,munoz13}. Using direct measurements of polar field, \citet{svalgaard05} and \citet{schatten05} predicted that the solar activity during SC 24 would be the lowest in the last 100 years, which has turned out to be correct.

Due to the importance of the tilt angle of SG to the poloidal field, the relationships of the tilt angle of SG with the parameters of the next SC have also been studied extensively in recent years based on the average tilt angle.
\cite{dasiespuig10} discovered a positive correlation between $max(\overline{S}\times\overline{\alpha_{\omega,\lambda}})$ and $max(\overline{S})$ of the next SC, where $\overline{S}$ was the monthly mean sunspot area from \cite{balmaceda09},
\begin{equation}
\overline{\alpha_{\omega,\lambda}}= \frac{\Sigma (\omega_{\text{j}} \times \alpha_{\text{j}} \times e^{-|\frac{\lambda_{\text{j}}}{10}|})}{\Sigma \omega_{\text{j}}},
\end{equation}
where $\omega_{\text{j}}$, $\alpha_{\text{j}}$, and $\lambda_{\text{j}}$ were the area, tilt angle, and latitude of the SG $j$ in a given month, respectively. The tilt coefficient does not show a clear correlation with the amplitude of the next SC \citep{tlatova18}.

One widely accepted generation mechanism responsible for the tilt angle of SG is the action of the Coriolis force on toroidally oriented flux tubes as they rise buoyantly through the convective zone and emerge at the photosphere \citep{wang91,dsilva93,fisher00}. There is a large scatter of the tilt angle of SG around the mean, which is thought to be due to the turbulent nature of the solar convection zone \citep{howard91,longcope02,bhowmik18,jiao21}. Numerical simulations have suggested that the component introduced by the tilt-angle scatter has a significant impact on the strength of the next SC \citep{baumann04,jiang14,bhowmik18}.

\citet{wang89} divided SGs into three types according to their fluxes: strong, medium, and weak SGs, then investigated their relationships of the tilt angle with latitude, respectively. They found that the average tilt angles of strong, medium, and weak SGs were essentially the same, and their values as a function of latitude (Joy's law and slope) were similar. However, the root-mean-square tilt angles of strong SGs differ from those of weak SGs having higher root-mean-square tilt angles (essentially having a higher standard deviation in the distribution). The root-mean-square value and the standard deviation are two statistical measures of the dispersion or spread of a group of data.

During SCs 21 and 22, the main trends over time of the averaged absolute values of tilt angles (simply the root-mean-square values) are similar to those of the average tilt angles, while, during SC 23, the main trend over time of the averaged absolute values of tilt angles is quite different from that of the average tilt angles \citep{gao23}. In addition, during the declining phase of SC 23, the absolute values of tilt angles are abnormally large \citep{gao23}. As we know, some of the most important indicators of solar activity during SC 24 reach their lowest levels in nearly a century.

The tilt angle of SG is crucial for the reversal of the polar field and the build-up of the opposite-polarity polar field. The low-latitude magnetic flux is transported to the pole taking about 2 years \citep{jiang14}. Thus, the tilt angle of SG during the declining phase (after the polarity reversal) is more crucial for the build-up of the opposite-polarity polar field, which is the source of the toroidal field responsible for solar activity during the next SC.

To better understand the operational mechanism of the solar dynamo, we investigate the relationships of the tilt angle of SG, including the mean tilt angle and the tilt-angle scatter, during the declining phase with the parameters of the next SC.

\section{Data} \label{sec:floats}

The data used in our analysis are derived from the white-light images taken at the KK \footnote{ftp://ftp.ngdc.noaa.gov/STP/SOLAR\_DATA/SUNSPOT\_REGIONS/Kodaikanal\_Tilt/}
and MW \footnote{ftp://ftp.ngdc.noaa.gov/STP/SOLAR\_DATA/SUNSPOT\_REGIONS/Mt\_Wilson\_Tilt/} observatories. The KK database provides a nearly continuous time series from 1906 to 1987, covering the whole extent of SCs 15 to 21, as well as the maximum and declining phases of SC 14. The MW database provides a nearly continuous time series from 1917 to 1985, covering the whole extent of SCs 16 to 20, almost the whole extent of SC 21, and the maximum and declining phases of SC 15. They both give comprehensive details, such as the tilt angle, latitude, area, and so on, for each SG event.

For the KK and MW data, the tilt angle of SG is defined using the following method \citep{howard84,howard91,sivaraman93,dasiespuig10}: 1. Group the sunspots: given a box, $3^\circ$ wide in latitude and $5^\circ$ wide in longitude, centered at each sunspot, any other sunspot is included as part of the SG if it falls inside the box. 2. Define the portion to the east of the mass center of the SG as the leading sunspots and the portion to the west as the following sunspots. 3. Calculate the tilt angle using the following formula:
\begin{equation}
\tan \alpha = \bigtriangleup \lambda / ( \cos \lambda \times \bigtriangleup l ),
\end{equation}
where $\bigtriangleup \lambda$ and $\bigtriangleup l$ represent the differences in latitudes and central meridian distances of the leading and following sunspots, respectively.

In the KK and MW databases, the tilt angles vary between [-90$^\circ$, 90$^\circ$]. The tilt angle is calculated based on the geometrical positions of the leading and following sunspots of a bipolar SG. Thus, the anti-Hale SGs can not be removed from the KK and MW databases. However, at present, they provide the longest available records of tilt angles. The published estimations of the percentage of anti-Hale SGs vary within a certain range: 2.4\% (Hale \& Nicholson 1925 found that there were 41 anti-Hale SGs out of 1735 SGs), 3.0\%\citep{zhukova20}, 3.1\%\citep{richardson48}, 4\%\citep{wang89}, 4.9\%\citep{khlystova09}, less than 5\% \citep{smith68}, 8.1\% \citep{li18}, and 8.4\% \citep{mcclintock14}. The results obtained from the widely used tilt-angle data in the KK and MW databases should show the general relationships of the tilt angle during the declining phase with the parameters of the next SC.

Following what \cite{dasiespuig10} did, we remove the data that the tilt angle is zero and the distance between the leading and following sunspots is larger than $16^\circ$.

For a SG, the KK and MW databases provide two (umbral) areas on days 1 and 2. However, they provide one tilt angle and one latitude on day 1. Thus, in this study, we use the area on day 1.

\section{Method}
Based on the KK data, Figure 1 shows the histograms of tilt-angle distributions for the declining phases of SCs 14-21, respectively. The solid lines in Figure 1 are the respective normal fits. The tilt-angle distribution during the declining phase is close to a normal distribution.

\begin{figure*}
   \centering
   \includegraphics[bb=25 175 570 685, width=0.8\hsize]{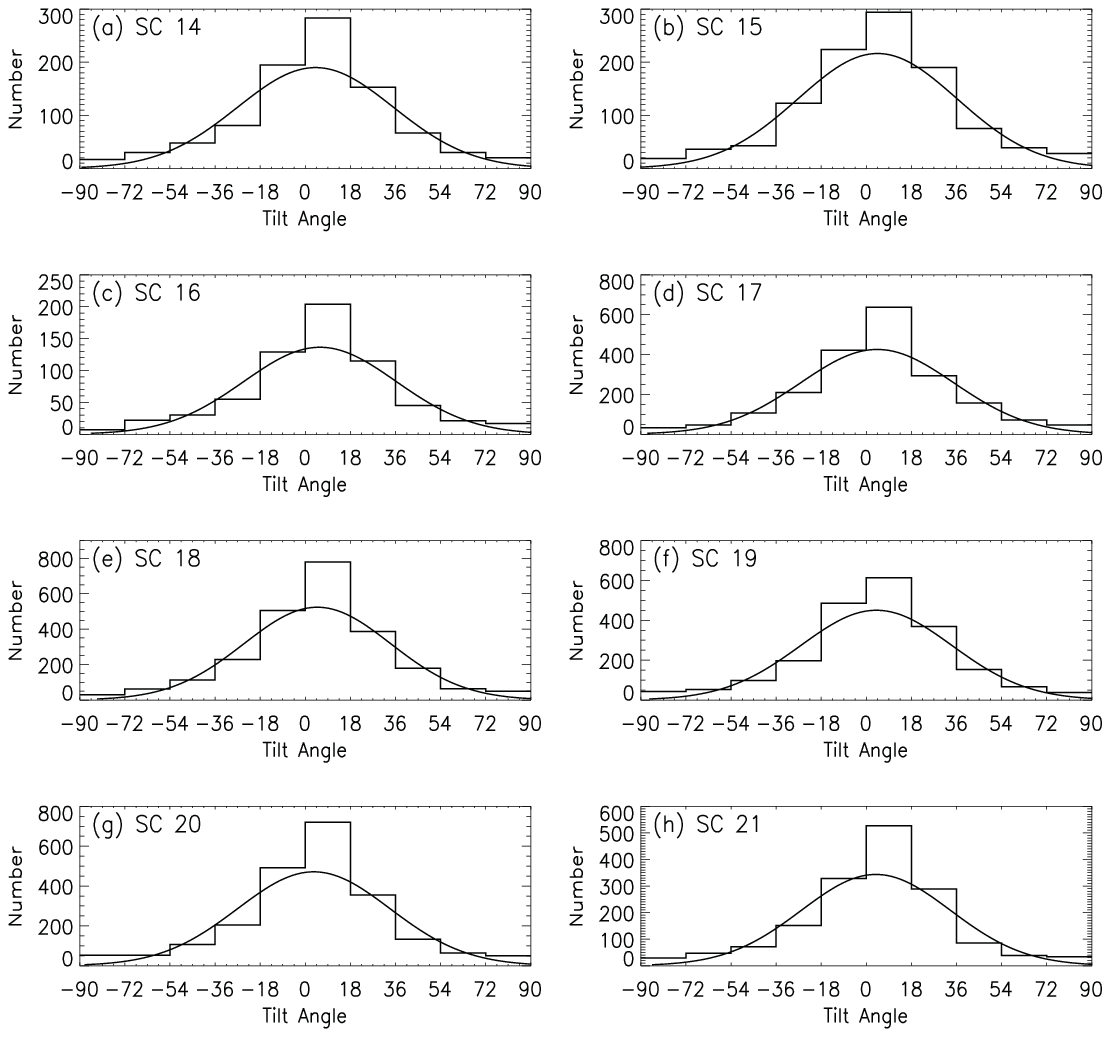}
      \caption{The tilt-angle distributions for the declining phases of SCs 14-21. The solid lines show the respective normal fits. These data are derived from the KK database.}
         \label{FigVibStab}
   \end{figure*}

Thus, for the tilt angles of SGs during the declining phase of a given SC: $\alpha_{\text{1}}$, $\alpha_{\text{2}}$, ..., $\alpha_{\text{n}}$, their mean is
\begin{equation}
\langle\alpha\rangle=\frac{\alpha_{\text{1}}+\alpha_{\text{2}}+...+\alpha_{\text{n}}}{n},
\end{equation}
their standard deviation is
\begin{equation}
\sigma=\sqrt\frac{({\alpha_{\text{1}}}-\langle\alpha\rangle)^2+({\alpha_{\text{2}}}-\langle\alpha\rangle)^2+...+(\alpha_{\text{n}}-\langle\alpha\rangle)^2}{n},
\end{equation}
and their root mean square is
\begin{equation}
RMS=\sqrt\frac{{\alpha_{\text{1}}}^2+{\alpha_{\text{2}}}^2+...+{\alpha_{\text{n}}}^2}{n}.
\end{equation}
The standard deviation is the root mean square deviation of values from their mean. If their mean is 0, their standard deviation is same as their root mean square. In this article, during the declining phase of a given SC, we calculate the standard deviation and the root mean square of tilt angles, which are two statistical measures of the dispersion or spread of data. We also calculate the mean tilt angle.

To take the area and latitude of SG into account in the relationships between the tilt angle during the declining phase and the parameters of the next SC, we calculate the area-weighted tilt angle,
\begin{equation}
\alpha_\omega = \frac{\sum (\omega_{\text{j}} \times \alpha_{\text{j}})}{\sum \omega_{\text{j}}},
\end{equation}
the latitude-weighted tilt angle,
\begin{equation}
\alpha_\lambda = \frac{\sum (|\lambda_{\text{j}}| \times \alpha_{\text{j}})}{\sum |\lambda_{\text{j}}|},
\end{equation}
and the area- and latitude- weighted tilt angle,
\begin{equation}
\alpha_{\omega,\lambda} = \frac{\sum (\omega_{\text{j}} \times |\lambda_{\text{j}}| \times \alpha_{\text{j}})}{\sum (\omega_{\text{j}} \times |\lambda_{\text{j}}|) }.
\end{equation}

Similarly, we also calculate the mean absolute value of tilt angles (simply the root mean square),
\begin{equation}
\langle|\alpha|\rangle=\frac{|\alpha_{\text{1}}|+|\alpha_{\text{2}}|+...+|\alpha_{\text{n}}|}{n},
\end{equation}
the area-weighted absolute value of tilt angles,
\begin{equation}
|\alpha_\omega|  = \frac{\sum (\omega_{\text{j}} \times |\alpha_{\text{j}}|)}{\sum \omega_{\text{j}}},
\end{equation}
the latitude-weighted absolute value of tilt angles,
\begin{equation}
|\alpha_\lambda|  = \frac{\sum |\lambda_{\text{j}}| \times |\alpha_{\text{j}}|}{\sum |\lambda_{\text{j}}|},
\end{equation}
and the area- and latitude- weighted absolute value of tilt angles,
\begin{equation}
|\alpha_{\omega,\lambda}| = \frac{\sum (\omega_{\text{j}} \times |\lambda_{\text{j}}| \times |\alpha_{\text{j}}|)}{\sum (\omega_{\text{j}} \times |\lambda_{\text{j}}|) }.
\end{equation}

We focus on three SC properties: the strength, the amplitude, and the duration of the ascending phase. \citet{balmaceda09} compiled a time series of sunspot area by combining different data sets (Royal Greenwich Observatory, Russian, Solar Optical Observing Network, Rome, Yunnan, and Catania), which have fewer data gaps. In this study, following the method used in \cite{dasiespuig10}, the strength of a SC is the total area of all sunspots during the SC, which is derived from the time series of sunspot area.

The amplitude of SC is the largest smoothed monthly mean number of sunspots, and the duration of the ascending phase is the time from the solar minimum to maximum, which are taken from the National Geophysical Data Center (NGDC) \footnote{ftp://ftp.ngdc.noaa.gov/STP/space-weather/solar-data/solar-indices/sunspot-numbers/cycle-data/}.

There is an overlap of 2-4 years between a given SC and the next SC in the butterfly diagram: after the solar minimum, the SGs belonging to the previous SC still emerge near the equator; before the solar minimum, the SGs belonging to the next SC already emerge at higher latitudes \citep{wilson87,li01,jiao21,gao23}. During the overlap, the tilt angles of SGs belonging to the previous SC are crucial for the build-up of the polar field, while the tilt angles of SGs belonging to the next SC tend to produce contributions to the opposite-polarity polar field. Additionally, there is a slight difference between the reversal time of polar field \citep{jiang13} and the time of solar maximum taken from NGDC.

To more accurately grasp the relationships of the tilt angle during the declining phase with the parameters of the next SC, in this study, the declining phase of a given SC is from $T_{\text{max}}+1$ to $T_{\text{min}}^{'}-2$, where $T_{\text{max}}$ is the time of solar maximum of the SC, $T_{\text{min}}^{'}$ is the time of solar minimum between the SC and the next SC, and the unit of $T_{\text{max}}$ and $T_{\text{min}}^{'}$ is the year.

Take SC 21 as an example, the solar maximum occurs in November 1979 (1979.9) and the solar minimum between SCs 21 and 22 occurs in October 1986 (1986.8) taken from NGDC. In this paper, the declining phase of SC 21 is from December 1980 to September 1984.

\section{Results} \label{sec:displaymath}

\subsection{The KK data}

\begin{figure*}
   \centering
   \includegraphics[bb=130 375 550 755, width=0.4\hsize]{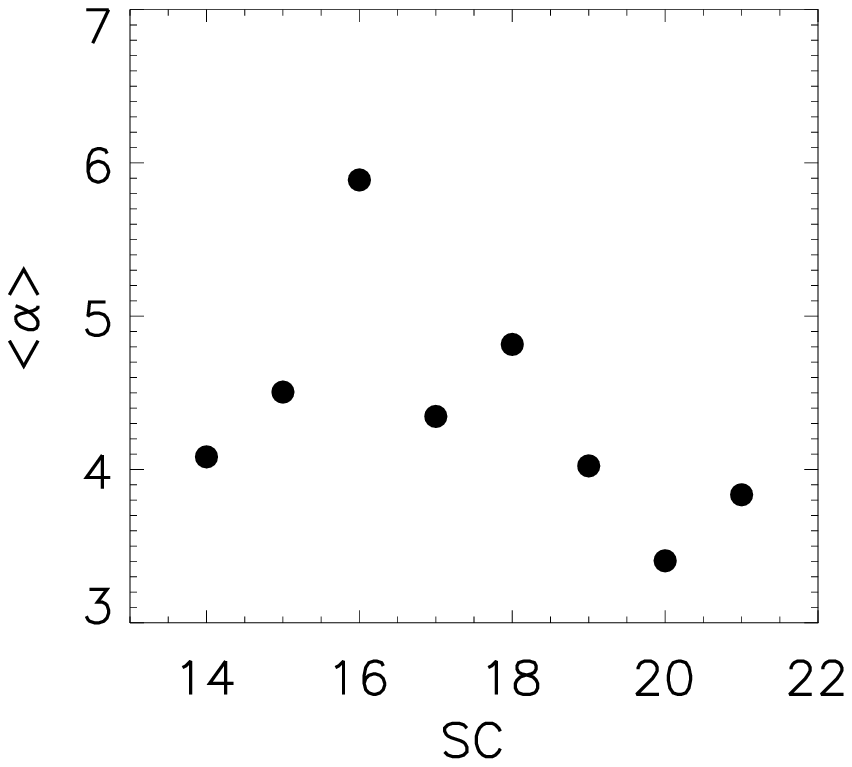}
      \caption{The mean tilt angles during the declining phases of SCs 14-21. These data are derived from the KK database.}
         \label{FigVibStab}
   \end{figure*}

\begin{figure*}
   \centering
   \includegraphics[bb=10 585 580 750, width=0.99\hsize]{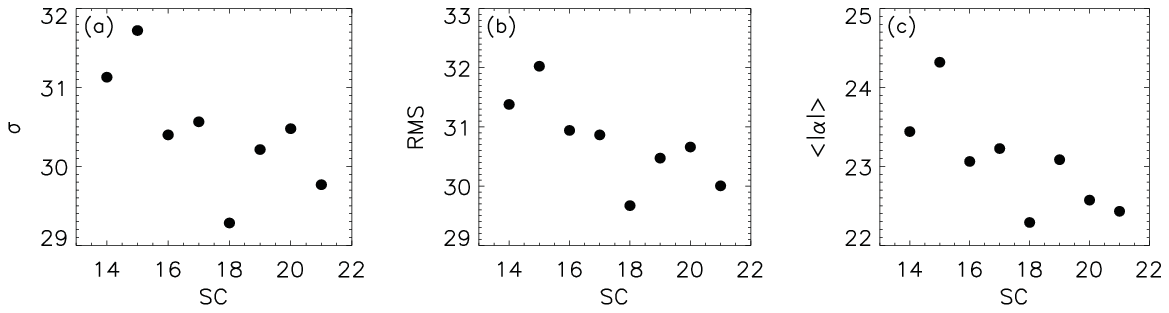}
      \caption{The standard deviations of tilt angles (a), the root-mean-square tilt angles (b), and the mean absolute values of tilt angles (c) during the declining phases of SCs 14-21. These data are derived from the KK database.}
         \label{FigVibStab}
   \end{figure*}

Figure 2 shows the mean tilt angles during the declining phases of SCs 14-21. Figure 3 shows the standard deviations of tilt angles, the root-mean-square tilt angles, and the mean absolute values of tilt angles during the declining phases of SCs 14-21. These data are derived from the KK database. Overall, the standard deviation of tilt angles, the root-mean-square tilt angle, and the mean absolute value of tilt angles increase or decrease synchronously.

\begin{figure*}
   \centering
   \includegraphics[bb=10 585 580 745, width=0.99\hsize]{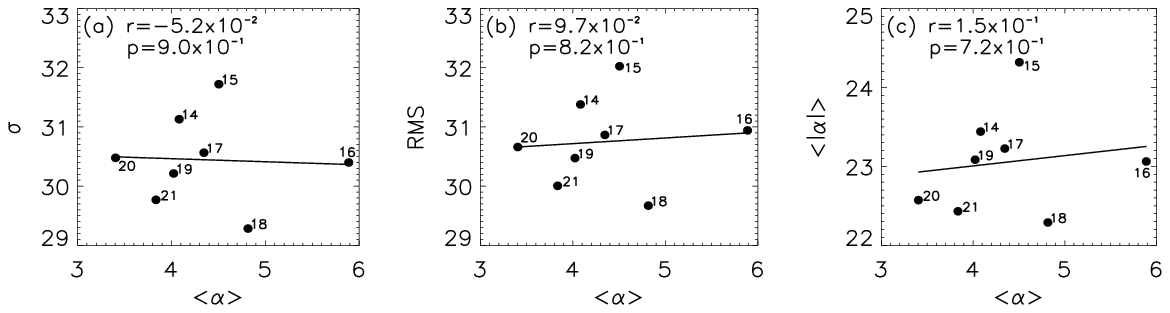}
      \caption{The standard deviation of tilt angles (a), the root-mean-square tilt angle (b), and the mean absolute value of tilt angles (c) vs. the mean tilt angle during the declining phase. The solid lines show the respective linear fits. The correlation coefficients and the probabilities that the correlations are due to chance can be found in each panel. These data are derived from the KK database.}
         \label{FigVibStab}
   \end{figure*}

\begin{table*}
 \centering
 \begin{minipage}{1.0\textwidth}
  \caption{The correlation coefficients between the mean tilt angle and the tilt-angle scatter (the standard deviation of tilt angles, the root-mean-square tilt angle, and the mean absolute value of tilt angles). These data are derived from the KK database.}
  \begin{tabular}{ccccccc}
  \hline
  \hline
Parameter & \multicolumn{2}{c}{$\sigma$ } & \multicolumn{2}{c}{RMS} & \multicolumn{2}{c}{$\langle|\alpha|\rangle$}\\
          & r & p & r & p & r & p\\
\hline
$\langle\alpha\rangle$ & $-5.2\times10^{-2}$ & $9.0\times10^{-1}$ & $9.7\times10^{-2}$ & $8.2\times10^{-1}$ & $1.5\times10^{-1}$ & $7.2\times10^{-1}$ \\
\hline
\hline
\end{tabular}
\end{minipage}
\end{table*}

However, the increases or decreases of the mean tilt angle and the tilt-angle scatter (the standard deviation of tilt angles, the root-mean-square tilt angle, and the mean absolute value of tilt angles) are not synchronous. Then, we calculate the correlation coefficients between the mean tilt angle and the tilt-angle scatter, which are shown in Figure 4 and Table 1. Considering the low number of SCs (8), Figure 4 and Table 1 also show the probabilities that the correlations are due to chance. If the probability is less than or equal to $1.0\times10^{-1}$, the correlation is considered statistically significant and not due to chance, whereas if the probability is larger than $1.0\times10^{-1}$, the correlation is statistically insignificant \citep{li09}. From Figure 4 and Table 1, we find that, during the declining phase, the correlation between the mean tilt angle and the tilt-angle scatter is statistically insignificant.

\begin{figure*}
   \centering
   \includegraphics[bb=10 585 580 745, width=0.99\hsize]{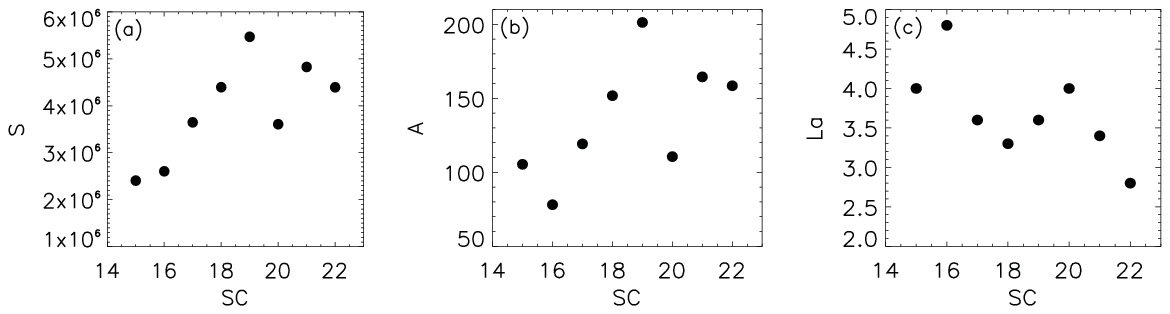}
      \caption{The strengths (a), amplitudes (b), and durations of the ascending phases (c) in SCs 15-22.}
         \label{FigVibStab}
   \end{figure*}

Figure 5 shows the strengths, amplitudes, and durations of the ascending phases in SCs 15-22. From Figures 4 and 5, we note that (1) the minimum value of the tilt-angle scatter occurs during the declining phase of SC 18, while the maximum values of the strength and amplitude all occur during SC 19; (2) the maximum value of the tilt-angle scatter occurs during the declining phase of SC 15, while the minimum value of the amplitude occurs during SC 16, and the strength of SC 16 is the second lowest.

\subsubsection{The relationships of the tilt angle during the declining phase with the parameters of the next SC.}

\begin{figure*}
   \centering
   \includegraphics[bb=10 95 585 750, width=0.99\hsize]{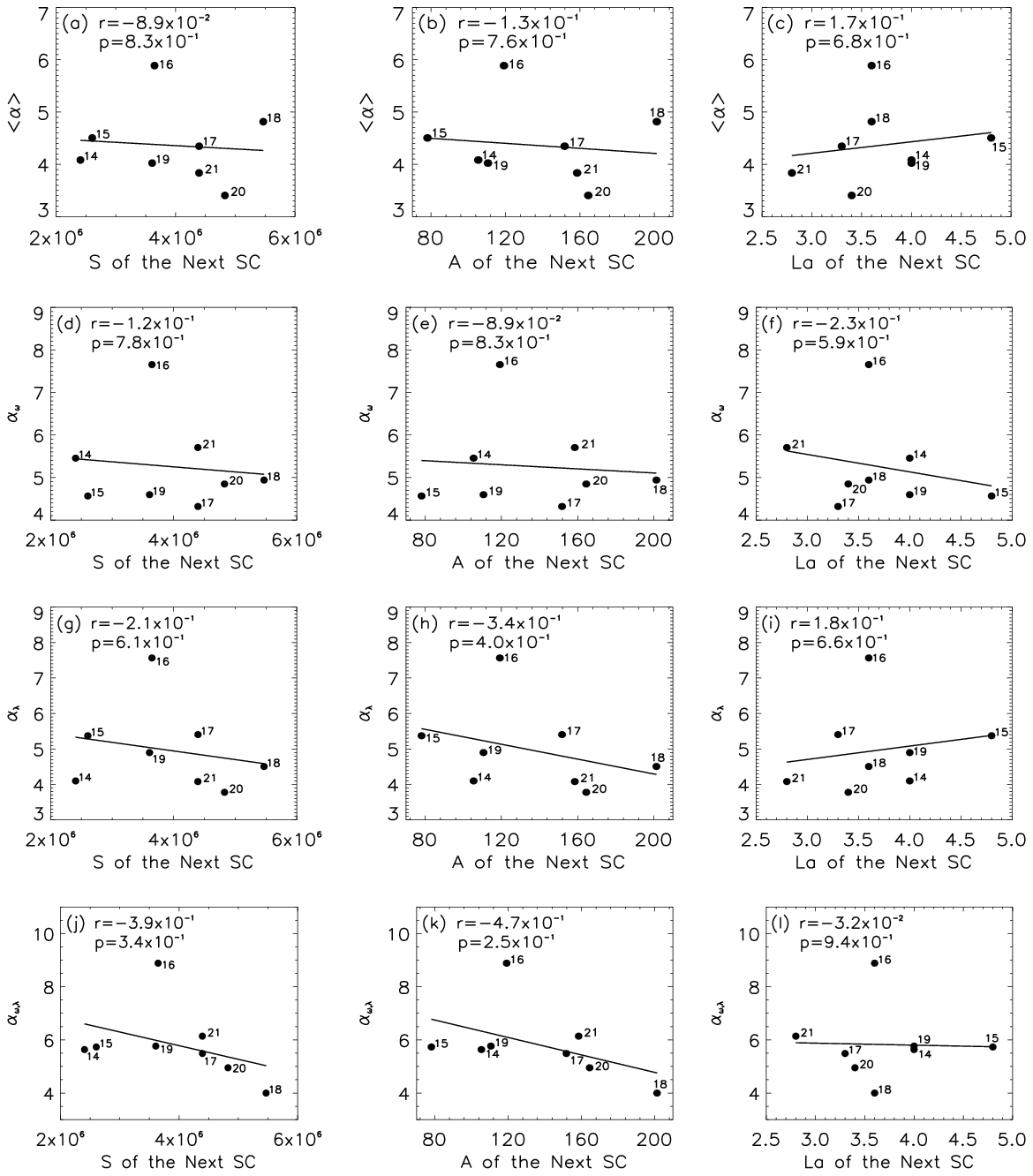}
      \caption{Panels (a), (b), and (c): the mean tilt angle during the declining phase vs. the strength, amplitude, and duration of the ascending phase of the next SC. Panels (d), (e), and (f): the area-weighted tilt angle during the declining phase vs. the strength, amplitude, and duration of the ascending phase of the next SC. Panels (g), (h), and (i): the latitude-weighted tilt angle during the declining phase vs. the strength, amplitude, and duration of the ascending phase of the next SC. Panels (j), (k), and (l): the area- and latitude- weighted tilt angle during the declining phase vs. the strength, amplitude, and duration of the ascending phase of the next SC. The solid lines show the respective linear fits. The correlation coefficients and probabilities can be found in each panel. These data are derived from the KK database.}
         \label{FigVibStab}
   \end{figure*}

\begin{table*}
 \centering
 \begin{minipage}{1.0\textwidth}
  \caption{The correlation coefficients of the 10 quantities based on the tilt angles during the declining phase with the strength(S), amplitude(A), and duration of the ascending phase (La) of the next SC. These data are derived from the KK database.}
  \begin{tabular}{ccccccc}
  \hline
  \hline
Parameter & \multicolumn{2}{c}{S} & \multicolumn{2}{c}{A} & \multicolumn{2}{c}{La}\\
          & r & p & r & p & r & p\\
\hline
$\langle\alpha\rangle$                                      & $-8.9\times10^{-2}$ & $8.3\times10^{-1}$ & $-1.3\times10^{-1}$ & $7.6\times10^{-1}$ & $1.7\times10^{-1}$  & $6.8\times10^{-1}$ \\
$\alpha_\omega$                                             & $-1.2\times10^{-1}$ & $7.8\times10^{-1}$ & $-8.9\times10^{-2}$ & $8.3\times10^{-1}$ & $-2.3\times10^{-1}$ & $5.9\times10^{-1}$ \\
$\alpha_\lambda$                                            & $-2.1\times10^{-1}$ & $6.1\times10^{-1}$ & $-3.4\times10^{-1}$ & $4.0\times10^{-1}$ & $1.8\times10^{-1}$  & $6.6\times10^{-1}$ \\
$\alpha_{\omega,\lambda}$                                   & $-3.9\times10^{-1}$ & $3.4\times10^{-1}$ & $-4.7\times10^{-1}$ & $2.5\times10^{-1}$ & $-3.2\times10^{-2}$ & $9.4\times10^{-1}$ \\
\hline
$\langle\alpha\rangle$ (without the data point of SC 16)    & $-7.5\times10^{-3}$ & $9.9\times10^{-1}$ & $2.6\times10^{-2}$  & $9.6\times10^{-1}$ & $3.9\times10^{-1}$  & $3.9\times10^{-1}$ \\
$\alpha_\omega$ (without the data point of SC 16)           & $-5.6\times10^{-2}$ & $9.1\times10^{-1}$ & $1.6\times10^{-1}$  & $7.4\times10^{-1}$ & $-4.0\times10^{-1}$ & $3.7\times10^{-1}$ \\
$\alpha_\lambda$ (without the data point of SC 16)          & $-2.5\times10^{-1}$ & $5.9\times10^{-1}$ & $-4.0\times10^{-1}$ & $3.8\times10^{-1}$ & $4.8\times10^{-1}$  & $2.8\times10^{-1}$ \\
$\alpha_{\omega,\lambda}$ (without the data point of SC 16) & $-6.4\times10^{-1}$ & $1.2\times10^{-1}$ & $-6.8\times10^{-1}$ & $9.5\times10^{-2}$ & $4.4\times10^{-2}$  & $9.3\times10^{-1}$ \\
\hline
$\sigma$                                                    & $-8.3\times10^{-1}$ & $1.0\times10^{-2}$ & $-8.4\times10^{-1}$ & $8.7\times10^{-3}$ & $7.1\times10^{-1}$  & $5.0\times10^{-2}$ \\
RMS                                                         & $-8.4\times10^{-1}$ & $8.8\times10^{-3}$ & $-8.6\times10^{-1}$ & $6.6\times10^{-3}$ & $7.3\times10^{-1}$  & $4.1\times10^{-2}$ \\
$\langle|\alpha|\rangle$                                    & $-8.5\times10^{-1}$ & $8.2\times10^{-3}$ & $-8.9\times10^{-1}$ & $3.4\times10^{-3}$ & $8.2\times10^{-1}$  & $1.3\times10^{-2}$ \\
$|\alpha_\omega|$                                           & $-7.8\times10^{-1}$ & $2.4\times10^{-2}$ & $-8.3\times10^{-1}$ & $9.9\times10^{-3}$ & $7.4\times10^{-1}$  & $3.7\times10^{-2}$ \\
$|\alpha_\lambda|$                                          & $-6.4\times10^{-1}$ & $8.8\times10^{-2}$ & $-7.5\times10^{-1}$ & $3.1\times10^{-2}$ & $7.0\times10^{-1}$  & $5.1\times10^{-2}$ \\
$|\alpha_{\omega,\lambda}|$                                 & $-7.7\times10^{-1}$ & $2.4\times10^{-2}$ & $-8.3\times10^{-1}$ & $1.1\times10^{-2}$ & $5.4\times10^{-1}$  & $1.7\times10^{-1}$ \\
\hline
\hline
\end{tabular}
\end{minipage}
\end{table*}

We investigate the relationships of the tilt angle, including the mean tilt angle and the tilt-angle scatter, during the declining phase with the strength and amplitude of the next SC. Figure 6 shows the mean tilt angle during the declining phase versus the strength and amplitude of the next SC. The correlation coefficients and probabilities are presented in Figure 6 and Table 2. Considering that SC 16 is an outlier, the correlation coefficients and probabilities are calculated without the data point of SC 16 (Table 2). With and without the data point of SC 16, the correlations of the mean tilt angle during the declining phase with the strength and amplitude of the next SC are statistically insignificant.

   \begin{figure*}
   \centering
   \includegraphics[bb=10 260 585 750, width=0.99\hsize]{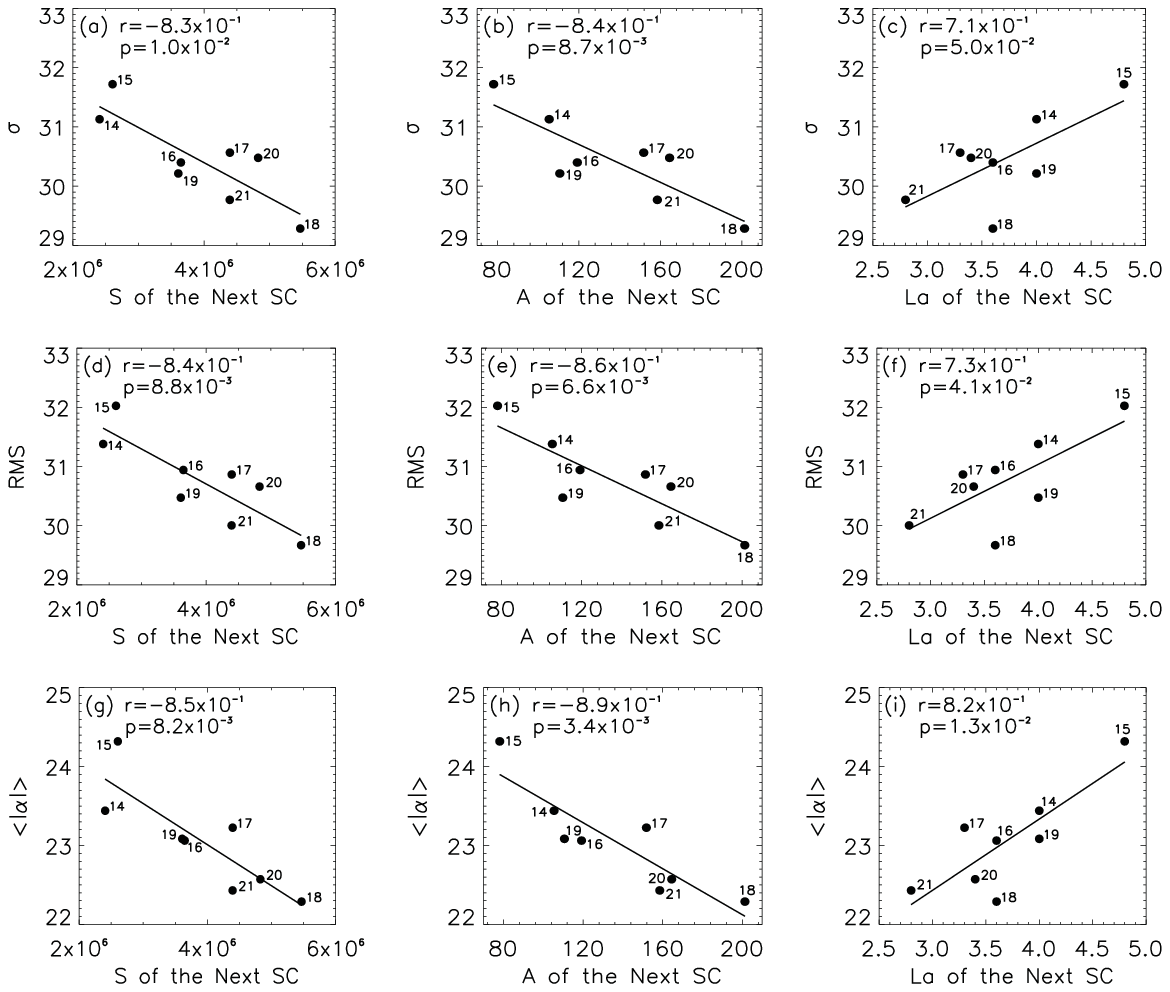}
      \caption{Panels (a), (b), and (c): the standard deviation of tilt angles during the declining phase vs. the strength, amplitude, and duration of the ascending phase of the next SC. Panels (d), (e), and (f): the root-mean-square tilt angle during the declining phase vs. the strength, amplitude, and duration of the ascending phase of the next SC. Panels (g), (h), and (i): the mean absolute value of tilt angles during the declining phase vs. the strength, amplitude, and duration of the ascending phase of the next SC. The solid lines show the respective linear fits. The correlation coefficients and probabilities can be found in each panel. These data are derived from the KK database.}
         \label{FigVibStab}
   \end{figure*}

Figure 7 shows the tilt-angle scatter during the declining phase versus the strength and amplitude of the next SC. The correlation coefficients and probabilities are shown in Figure 7 and Table 2. The standard deviation of tilt angles during the declining phase shows significant anti-correlations with the strength and amplitude of the next SC, with correlation coefficients of $r = -8.3\times10^{-1}$ at 99.0\% confidence level and $ r = -8.4\times10^{-1}$ at 99.13\% confidence level, respectively; the root-mean-square tilt angle during the declining phase exhibits significant anti-correlations with the strength and amplitude of the next SC, $ r = -8.4\times10^{-1}$ at 99.12\% confidence level and $ r = -8.6\times10^{-1}$ at 99.34\% confidence level, respectively; the mean absolute value of tilt angles during the declining phase has significant anti-correlations with the strength and amplitude of the next SC, $ r = -8.5\times10^{-1}$ at 99.18\% confidence level and $ r = -8.9\times10^{-1}$ at 99.66\% confidence level, respectively. The tilt-angle scatter during the declining phase has significant anti-correlations with the strength and amplitude of the next SC.

Next, we investigate the relationships of the mean tilt angle and the tilt-angle scatter during the declining phase with the duration of the ascending phase of the next SC (Figure 6, Figure 7, and Table 2). The positive correlations between the mean tilt angle during the declining phase and the duration of the ascending phase of the next SC are statistically insignificant with and without SC 16 (an outlier). \citet{dasiespuig10} found that the anti-correlation between the mean tilt angle during the whole extent of a SC and the length of the next SC is also statistically insignificant ($ P > 1.0\times10^{-1}$) based on the KK data.

While the tilt-angle scatter during the declining phase has a significant positive correlation with the duration of the ascending phase of the next SC: the standard deviation of tilt angles during the declining phase shows a significant positive correlation with the duration of the ascending phase of the next SC, with a correlation coefficient of $ r = 7.1\times10^{-1}$ at 95.0\% confidence level; the root-mean-square tilt angle during the declining phase exhibits a significant positive correlation with the duration of the ascending phase of the next SC, $ r = 7.3\times10^{-1}$ at 95.9\% confidence level; the mean absolute value of tilt angles during the declining phase has a significant positive correlation with the duration of the ascending phase of the next SC, $ r = 8.2\times10^{-1}$ at 98.7\% confidence level.

Figure 6 and Table 2 also show the relationships of the area-weighted tilt angle, the latitude-weighted tilt angle, and the area- and latitude- weighted tilt angle during the declining phase with the strength, amplitude, and duration of the ascending phase of the next SC. The correlations of these 3 quantities based on the tilt angles during the declining phase with the strength, amplitude, and duration of the ascending phase of the next SC are statistically insignificant with and without SC 16 (an outlier). An exception is the correlation of the area- and latitude- weighted tilt angle during the declining phase with the amplitude of the next SC without SC 16 ($ r = -6.8\times10^{-1}$ at 90.5\% confidence level). The anti-correlation between the area-weighted tilt angle during the whole extent of a SC and the length of the next SC is also statistically insignificant based on the KK data \citep{dasiespuig10}.

\begin{figure*}
   \centering
   \includegraphics[bb=10 260 585 750, width=0.99\hsize]{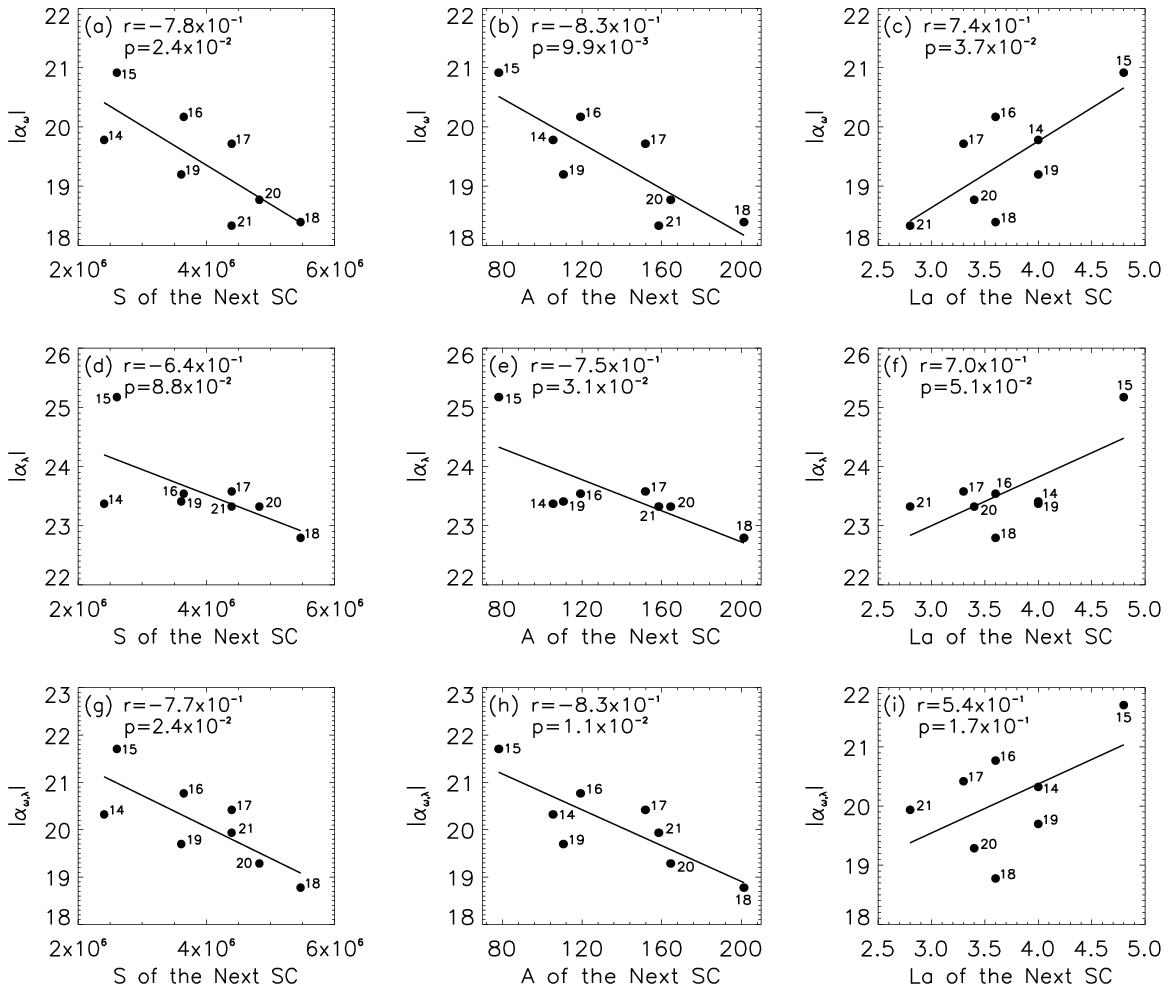}
      \caption{Panels (a), (b), and (c): the area-weighted absolute value of tilt angles during the declining phase vs. the strength, amplitude, and duration of the ascending phase of the next SC. Panels (d), (e), and (f): the latitude-weighted absolute value of tilt angles during the declining phase vs. the strength, amplitude, and duration of the ascending phase of the next SC. Panels (g), (h), and (i): the area- and latitude- weighted absolute value of tilt angles during the declining phase vs. the strength, amplitude, and duration of the ascending phase of the next SC. The solid lines show the respective linear fits. The correlation coefficients and probabilities can be found in each panel. These data are derived from the KK database.}
         \label{FigVibStab}
   \end{figure*}

We also investigate the relationships of the area-weighted absolute value of tilt angles, the latitude-weighted absolute value of tilt angles, and the area- and latitude- weighted absolute value of tilt angles during the declining phase with the strength, amplitude, and duration of the ascending phase of the next SC (Figure 8 and Table 2). The area-weighted absolute value of tilt angles during the declining phase has significant anti-correlations with the strength and amplitude of the next SC, $ r = -7.8\times10^{-1}$ at 97.6\% confidence level and $ r = -8.3\times10^{-1}$ at 99.01\% confidence level, respectively, and a positive correlation with the duration of the ascending phase of the next SC, $ r = 7.4\times10^{-1}$ at 96.3\% confidence level. The latitude-weighted absolute value of tilt angles during the declining phase exhibits significant anti-correlations with the strength and amplitude of the next SC, $ r = -6.4\times10^{-1}$ at 91.2\% confidence level and $ r = -7.5\times10^{-1}$ at 96.9\% confidence level, respectively, and a positive correlation with the duration of the ascending phase of the next SC, $ r = 7.0\times10^{-1}$ at 94.9\% confidence level. The area- and latitude- weighted absolute value of tilt angles during the declining phase shows significant anti-correlations with the strength and amplitude of the next SC, $ r = -7.7\times10^{-1}$ at 97.6\% confidence level and $ r = -8.3\times10^{-1}$ at 98.9\% confidence level, respectively.

\subsubsection{The relationships of the tilt-angle scatter with the statistical properties of SGs}

\citet{wang89} divided SGs into three types according to their fluxes: strong, medium, and weak SGs and there are 929 (34\%) weak SGs, 1071 (40\%) medium SGs, and 710 (26\%) strong SGs. To some extent, the area of SG acts as a proxy for the total magnetic flux \citep{zharkov06,dikpati06}. The tilt-angle scatter decreases with increasing area of SG \citep{wang89,jiang14}. To understand the effect of the distribution of the area (magnetic flux) on the tilt-angle scatter, we investigate the relationships of the tilt-angle scatter with the percentage of weak SGs, the percentage of medium SGs, the percentage of large SGs, the mean area of all SGs, and the total area of all SGs during the declining phase. We also investigate the relationship of the tilt-angle scatter with the number of all SGs during the declining phase.

We separate SGs into three categories as follows: weak SGs, $\omega < $ 9 millionths of the solar hemisphere (msh), medium SGs, 9 msh $\leq\omega \leq$ 27 msh, and strong SGs,  $\omega >$ 27 msh based on the KK data. During the declining phases of SCs 14-21, there are 4635 (36\%) weak SGs, 4988 (38\%) medium SGs, and 3377 (26\%) strong SGs.

\begin{figure*}
   \centering
   \includegraphics[bb=10 420 585 750, width=0.99\hsize]{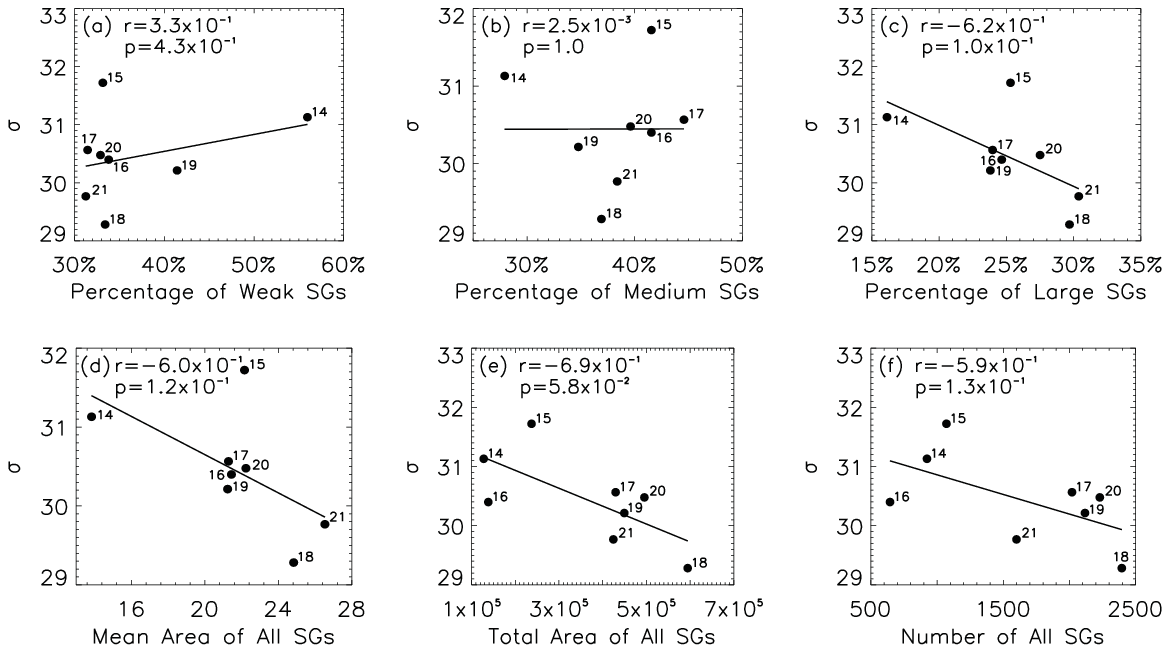}
      \caption{The standard deviation of tilt angles vs. the percentage of weak SGs (a), the percentage of medium SGs (b), the percentage of large SGs (c), the mean area of all SGs (d), the total area of all SGs (e), and the number of all SGs (f) during the declining phase. The solid lines show the respective linear fits. The correlation coefficients and probabilities can be found in each panel. These data are derived from the KK database.}
         \label{FigVibStab}
   \end{figure*}

\begin{table*}
 \centering
 \begin{minipage}{1.0\textwidth}
  \caption{The correlation coefficients of the standard deviation of tilt angles with the percentage of weak SGs, the percentage of medium SGs, the percentage of large SGs, the mean area of all SGs, the total area of all SGs, and the number of all SGs during the declining phase. These data are derived from the KK database.}
  \begin{tabular}{ccc}
  \hline
  \hline
Parameter & \multicolumn{2}{c}{$\sigma$} \\
          & r & p \\
\hline
Percentage of weak SGs                                     & $3.3\times10^{-1}$  & $4.3\times10^{-1}$ \\
Percentage of medium SGs                                   & $2.5\times10^{-3}$  & 1.0                \\
Percentage of large SGs                                    & $-6.2\times10^{-1}$ & $1.0\times10^{-1}$ \\
Mean area of all SGs                                       & $-6.0\times10^{-1}$ & $1.2\times10^{-1}$ \\
Total area of all SGs                                      & $-6.9\times10^{-1}$ & $5.8\times10^{-2}$ \\
Number of all SGs                                          & $-5.9\times10^{-1}$ & $1.3\times10^{-1}$ \\
\hline
Percentage of weak SGs (without the data point of SC 15)   & $6.1\times10^{-1}$  & $1.4\times10^{-1}$ \\
Percentage of medium SGs (without the data point of SC 15) & $-2.6\times10^{-1}$ & $5.8\times10^{-1}$ \\
Percentage of large SGs (without the data point of SC 15)  & $-8.6\times10^{-1}$ & $1.3\times10^{-2}$ \\
Mean area of all SGs (without the data point of SC 15)     & $-8.6\times10^{-1}$ & $1.2\times10^{-2}$ \\
Total area of all SGs (without the data point of SC 15)    & $-7.0\times10^{-1}$ & $7.7\times10^{-2}$ \\
Number of all SGs (without the data point of SC 15)        & $-5.2\times10^{-1}$ & $2.3\times10^{-1}$ \\
\hline
\hline
\end{tabular}
\end{minipage}
\end{table*}

Figure 9 shows the standard deviation of tilt angles versus the percentage of weak SGs, the percentage of medium SGs, the percentage of large SGs, the mean area of all SGs, the total area of all SGs, and the number of all SGs during the declining phase. The correlation coefficients and probabilities are shown in Figure 9 and Table 3. The standard deviation of tilt angles has significant anti-correlations with the percentage of large SGs and the total area of all SGs, $ r = -6.2\times10^{-1}$ at 90\% confidence level and $ r = -6.9\times10^{-1}$ at 94.2\% confidence level, respectively.

Considering that SC 15 is an outlier, the correlation coefficients and probabilities are calculated without the data point of SC 15 (Table 3). Without the data point of SC 15, the standard deviation of tilt angles has significant anti-correlations with the percentage of large SGs, the mean area of all SGs, and the total area of all SGs, with correlation coefficients of $ r = -8.6\times10^{-1}$ at 98.7\% confidence level, $ r = -8.6\times10^{-1}$ at 98.8\% confidence level, and $ r = -7.0\times10^{-1}$ at 92.3\% confidence level, respectively.

\subsubsection{The relationships of the tilt angle during the declining phase with the parameters of the same SC.}

\begin{table*}
 \centering
 \begin{minipage}{1.0\textwidth}
  \caption{The correlation coefficients of the 10 quantities based on the tilt angles during the declining phase with the strength (S), amplitude (A), and duration of the ascending phase (La) of the same SC. These data are derived from the KK database.}
  \begin{tabular}{ccccccc}
  \hline
  \hline
Parameter & \multicolumn{2}{c}{S} & \multicolumn{2}{c}{A} & \multicolumn{2}{c}{La}\\
          & r & p & r & p & r & p\\
\hline
$\langle\alpha\rangle$      & $-3.4\times10^{-1}$ & $4.1\times10^{-1}$ & $-3.2\times10^{-1}$ & $4.3\times10^{-1}$ & $2.7\times10^{-1}$ & $5.2\times10^{-1}$ \\
$\alpha_\omega$             & $-3.1\times10^{-1}$ & $4.5\times10^{-1}$ & $-4.1\times10^{-1}$ & $3.1\times10^{-1}$ & $5.2\times10^{-1}$ & $1.8\times10^{-1}$ \\
$\alpha_\lambda$            & $-3.2\times10^{-1}$ & $4.4\times10^{-1}$ & $-3.0\times10^{-1}$ & $4.8\times10^{-1}$ & $2.8\times10^{-1}$ & $5.1\times10^{-1}$ \\
$\alpha_{\omega,\lambda}$   & $-3.2\times10^{-1}$ & $4.4\times10^{-1}$ & $-3.5\times10^{-1}$ & $4.0\times10^{-1}$ & $5.0\times10^{-1}$ & $2.1\times10^{-1}$ \\
\hline
$\sigma$                    & $-7.3\times10^{-1}$ & $3.9\times10^{-2}$ & $-5.9\times10^{-1}$ & $1.3\times10^{-1}$ & $5.7\times10^{-1}$ & $1.4\times10^{-1}$ \\
RMS                         & $-7.8\times10^{-1}$ & $2.2\times10^{-2}$ & $-6.3\times10^{-1}$ & $9.1\times10^{-2}$ & $6.1\times10^{-1}$ & $1.1\times10^{-1}$ \\
$\langle|\alpha|\rangle$    & $-6.1\times10^{-1}$ & $1.0\times10^{-1}$ & $-4.1\times10^{-1}$ & $3.1\times10^{-1}$ & $4.1\times10^{-1}$ & $3.1\times10^{-1}$ \\
$|\alpha_\omega|$           & $-7.3\times10^{-1}$ & $3.9\times10^{-2}$ & $-5.7\times10^{-1}$ & $1.4\times10^{-1}$ & $5.4\times10^{-1}$ & $1.7\times10^{-1}$ \\
$|\alpha_\lambda|$          & $-4.7\times10^{-1}$ & $2.3\times10^{-1}$ & $-2.6\times10^{-1}$ & $5.4\times10^{-1}$ & $1.4\times10^{-1}$ & $7.5\times10^{-1}$ \\
$|\alpha_{\omega,\lambda}|$ & $-6.4\times10^{-1}$ & $9.0\times10^{-2}$ & $-4.8\times10^{-1}$ & $2.3\times10^{-1}$ & $4.2\times10^{-1}$ & $3.0\times10^{-1}$ \\
\hline
\hline
\end{tabular}
\end{minipage}
\end{table*}

Finally, we investigate the relationships of the tilt angle, including the mean tilt angle and the tilt-angle scatter, during the declining phase with the strength, amplitude, and duration of the ascending phase of the same SC.

The correlations of the mean tilt angle, the area-weighted tilt angle, the latitude-weighted tilt angle, and the area- and latitude- weighted tilt angle during the declining phase with the strength, amplitude, and duration of the ascending phase of the same SC are statistically insignificant (Table 4).

Five measures of dispersion during the declining phase exhibits significant anti-correlations with the strength of the same SC: the standard deviation of tilt angles ($r = -7.3\times10^{-1}$ at 96.1\% confidence level), the root-mean-square tilt angle ($r = -7.8\times10^{-1}$ at 97.8\% confidence level), the mean absolute value of tilt angles ($r = -6.1\times10^{-1}$ at 90\% confidence level), the area-weighted absolute value of tilt angles ($ r = -7.3\times10^{-1}$ at 96.1\% confidence level), and the area- and latitude- weighted absolute value of tilt angles ($r = -6.4\times10^{-1}$ at 91.0\% confidence level). The root-mean-square tilt angle during the declining phase shows a significant anti-correlation with the amplitude of the same SC ($r = -6.3\times10^{-1}$ at 90.9\% confidence level). The reason for these significant anti-correlations may be that the scatter of tilt angles has a significant anti-correlation with the total area of all SGs.

For the KK data, the anti-correlations of the mean tilt angle during the declining phase with the strength and amplitude of the same SC are statistically insignificant (Table 4), while the anti-correlations of the mean tilt angle during the whole extent of a SC with the strength and amplitude of the same SC are statistically significant \citep{dasiespuig10}. The anti-correlation between the mean tilt angle during the declining phase and the duration of the ascending phase of the same SC is statistically insignificant (Table 4). And the anti-correlation between the mean tilt angle during the whole extent of a SC and the length of the same SC is also statistically insignificant \citep{dasiespuig10}. The correlations of the area-weighted tilt angle during the declining phase with the strength, amplitude, and duration of the ascending phase of the same SC are statistically insignificant (Table 4). \citet{dasiespuig10} found that the correlations of the area-weighted tilt angle during the whole extent of a SC with the strength, amplitude, and length of the same SC are also statistically insignificant.

\subsection{The MW data}

To further validate the above findings, we utilize the data from the MW observatory and investigate the relationships of the tilt angle during the declining phase with the SC parameters. The results derived from the MW database generally corroborate those from the KK database. The correlations between the mean tilt angle and the tilt-angle scatter are statistically insignificant with and without SC 15 (an outlier) (Figure 10 and Table 5).

\begin{table*}
 \centering
 \begin{minipage}{1.0\textwidth}
  \caption{The correlation coefficients between the mean tilt angle and the tilt-angle scatter. These data are derived from the MW database.}
  \begin{tabular}{ccccccc}
  \hline
  \hline
Parameter & \multicolumn{2}{c}{$\sigma$ } & \multicolumn{2}{c}{RMS} & \multicolumn{2}{c}{$\langle|\alpha|\rangle$}\\
          & r & p & r & p & r & p\\
\hline
$\langle\alpha\rangle$                                   & $-3.7\times10^{-1}$ & $4.1\times10^{-1}$ & $-2.8\times10^{-1}$ & $5.4\times10^{-1}$ & $-1.8\times10^{-1}$ & $7.0\times10^{-1}$ \\
\hline
$\langle\alpha\rangle$ (without the data point of SC 15) & $-4.0\times10^{-1}$ & $4.3\times10^{-1}$ & $-2.1\times10^{-1}$ & $7.0\times10^{-1}$ & $1.4\times10^{-2}$  & $9.8\times10^{-1}$ \\
\hline
\end{tabular}
\end{minipage}
\end{table*}

\begin{figure*}
   \centering
   \includegraphics[bb=10 585 580 745, width=0.99\hsize]{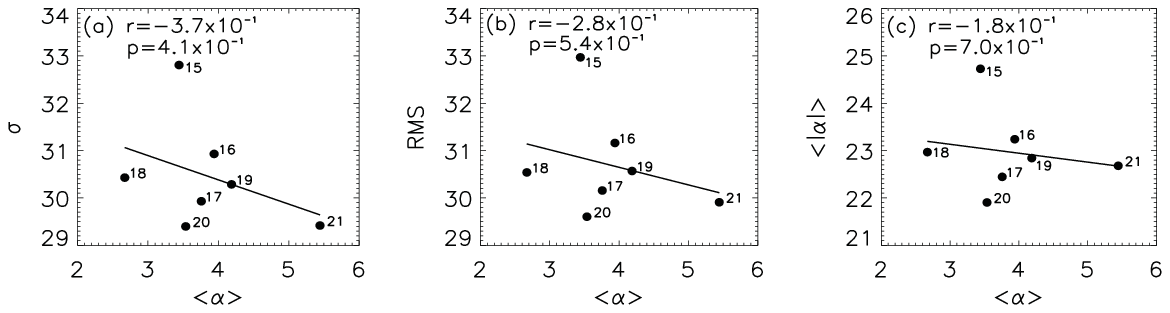}
      \caption{The standard deviation of tilt angles (a), the root-mean-square tilt angle (b), and the mean absolute value of tilt angles (c) vs. the mean tilt angle during the declining phase. The solid lines show the respective linear fits. The correlation coefficients and probabilities can be found in each panel. These data are derived from the MW database.}
         \label{FigVibStab}
   \end{figure*}

\subsubsection{The relationships of the tilt angle during the declining phase with the parameters of the next SC.}

\begin{figure*}
   \centering
   \includegraphics[bb=10 95 585 750, width=0.99\hsize]{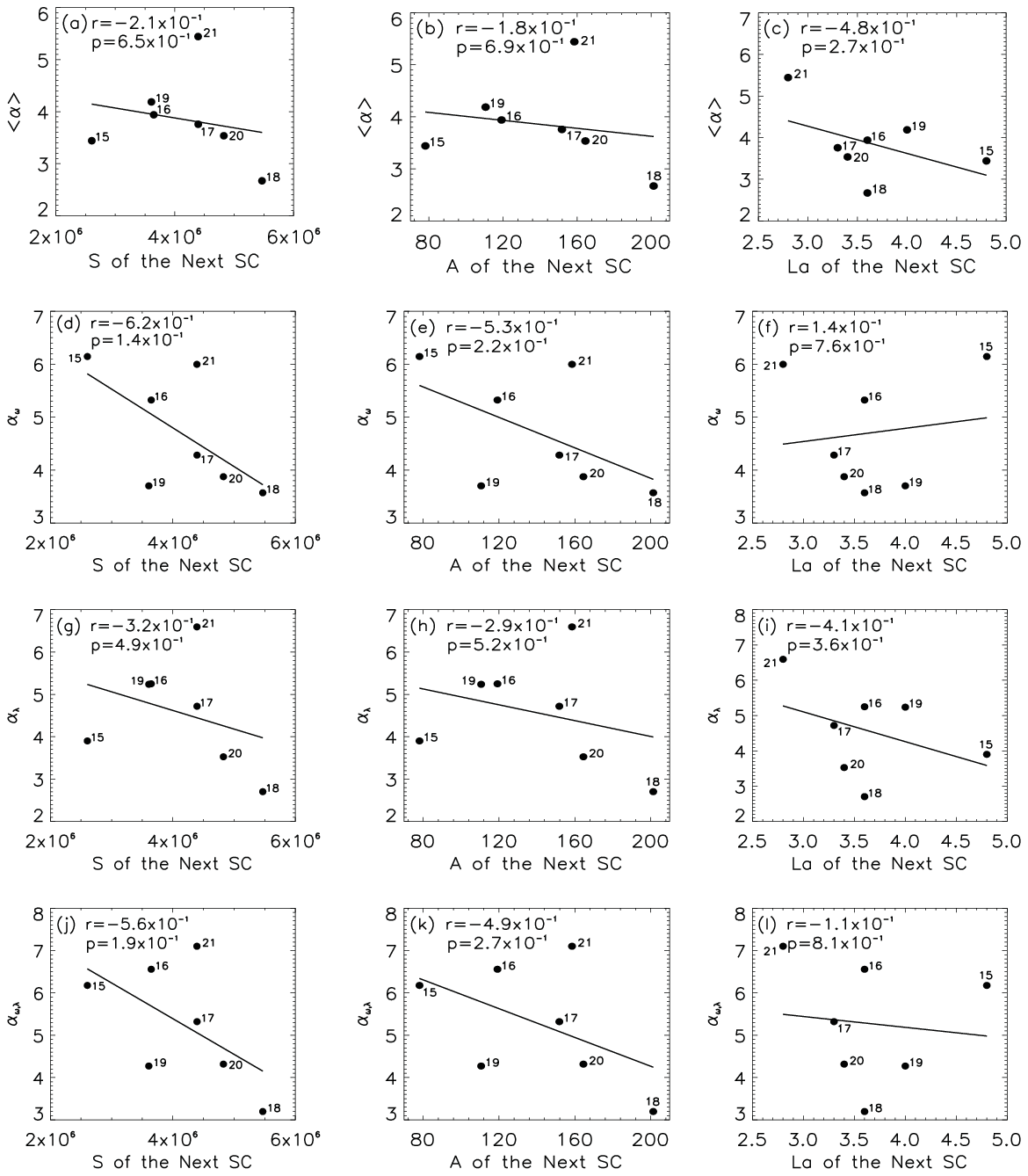}
      \caption{Panels (a), (b), and (c): the mean tilt angle during the declining phase vs. the strength, amplitude, and duration of the ascending phase of the next SC. Panels (d), (e), and (f): the area-weighted tilt angle during the declining phase vs. the strength, amplitude, and duration of the ascending phase of the next SC. Panels (g), (h), and (i): the latitude-weighted tilt angle during the declining phase vs. the strength, amplitude, and duration of the ascending phase of the next SC. Panels (j), (k), and (l): the area- and latitude- weighted tilt angle during the declining phase vs. the strength, amplitude, and duration of the ascending phase of the next SC. The solid lines show the respective linear fits. The correlation coefficients and probabilities can be found in each panel. These data are derived from the MW database.}
         \label{FigVibStab}
   \end{figure*}

\begin{table*}
 \centering
 \begin{minipage}{1.0\textwidth}
  \caption{The correlation coefficients of the 10 quantities based on the tilt angles during the declining phase with the strength (S), amplitude (A) and duration of the ascending phase (La) of the next SC. These data are derived from the MW database.}
  \begin{tabular}{ccccccc}
  \hline
  \hline
Parameter & \multicolumn{2}{c}{S} & \multicolumn{2}{c}{A} & \multicolumn{2}{c}{La}\\
          & r & p & r & p & r & p\\
\hline
$\langle\alpha\rangle$                 & $-2.1\times10^{-1}$ & $6.5\times10^{-1}$ & $-1.8\times10^{-1}$ & $6.9\times10^{-1}$ & $-4.8\times10^{-1}$ & $2.7\times10^{-1}$ \\
$\alpha_\omega$                        & $-6.2\times10^{-1}$ & $1.4\times10^{-1}$ & $-5.3\times10^{-1}$ & $2.2\times10^{-1}$ & $1.4\times10^{-1}$  & $7.6\times10^{-1}$ \\
$\alpha_\lambda$                       & $-3.2\times10^{-1}$ & $4.9\times10^{-1}$ & $-2.9\times10^{-1}$ & $5.2\times10^{-1}$ & $-4.1\times10^{-1}$ & $3.6\times10^{-1}$ \\
$\alpha_{\omega,\lambda}$              & $-5.6\times10^{-1}$ & $1.9\times10^{-1}$ & $-4.9\times10^{-1}$ & $2.7\times10^{-1}$ & $-1.1\times10^{-1}$ & $8.1\times10^{-1}$ \\
\hline
$\sigma$                               & $-7.5\times10^{-1}$ & $5.4\times10^{-2}$ & $-7.1\times10^{-1}$ & $7.4\times10^{-2}$ & $9.0\times10^{-1}$  & $6.3\times10^{-3}$ \\
RMS                                    & $-7.9\times10^{-1}$ & $3.6\times10^{-2}$ & $-7.4\times10^{-1}$ & $5.5\times10^{-2}$ & $8.7\times10^{-1}$  & $9.9\times10^{-3}$ \\
$\langle|\alpha|\rangle$               & $-7.5\times10^{-1}$ & $5.3\times10^{-2}$ & $-6.9\times10^{-1}$ & $8.8\times10^{-2}$ & $8.0\times10^{-1}$  & $3.0\times10^{-2}$ \\
$|\alpha_\omega|$                      & $-8.3\times10^{-1}$ & $2.2\times10^{-2}$ & $-8.0\times10^{-1}$ & $3.2\times10^{-2}$ & $8.7\times10^{-1}$  & $1.2\times10^{-2}$ \\
$|\alpha_\lambda|$                     & $-7.5\times10^{-1}$ & $5.2\times10^{-2}$ & $-7.4\times10^{-1}$ & $5.6\times10^{-2}$ & $9.5\times10^{-1}$  & $1.1\times10^{-3}$ \\
$|\alpha_{\omega,\lambda}|$            & $-8.2\times10^{-1}$ & $2.5\times10^{-2}$ & $-8.4\times10^{-1}$ & $1.9\times10^{-2}$ & $8.1\times10^{-1}$  & $2.9\times10^{-2}$ \\
\hline
\hline
\end{tabular}
\end{minipage}
\end{table*}

Similarly, the correlations of the mean tilt angle, the area-weighted tilt angle, the latitude-weighted tilt angle, and the area- and latitude- weighted tilt angle during the declining phase with the strength, amplitude, and duration of the ascending phase of the next SC are statistically insignificant (Figure 11 and Table 6).

\begin{figure*}
   \centering
   \includegraphics[bb=10 260 585 750, width=0.99\hsize]{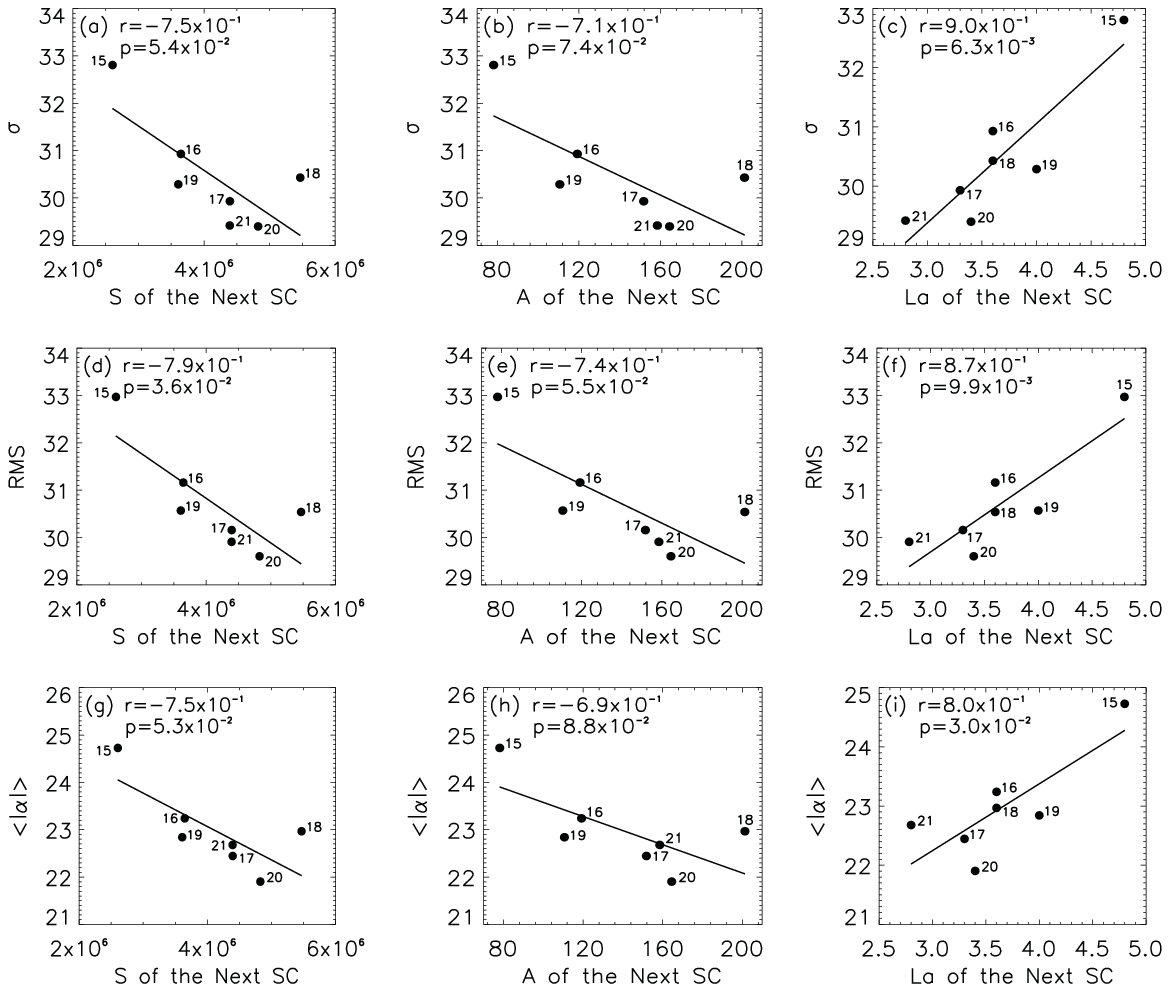}
      \caption{Panels (a), (b), and (c): the standard deviation of tilt angles during the declining phase vs. the strength, amplitude, and duration of the ascending phase of the next SC. Panels (d), (e), and (f): the root-mean-square tilt angle during the declining phase vs. the strength, amplitude, and duration of the ascending phase of the next SC. Panels (g), (h), and (i): the mean absolute value of tilt angles during the declining phase vs. the strength, amplitude, and duration of the ascending phase of the next SC. The solid lines show the respective linear fits. The correlation coefficients and probabilities can be found in each panel. These data are derived from the MW database.}
         \label{FigVibStab}
   \end{figure*}

\begin{figure*}
   \centering
   \includegraphics[bb=10 260 585 750, width=0.99\hsize]{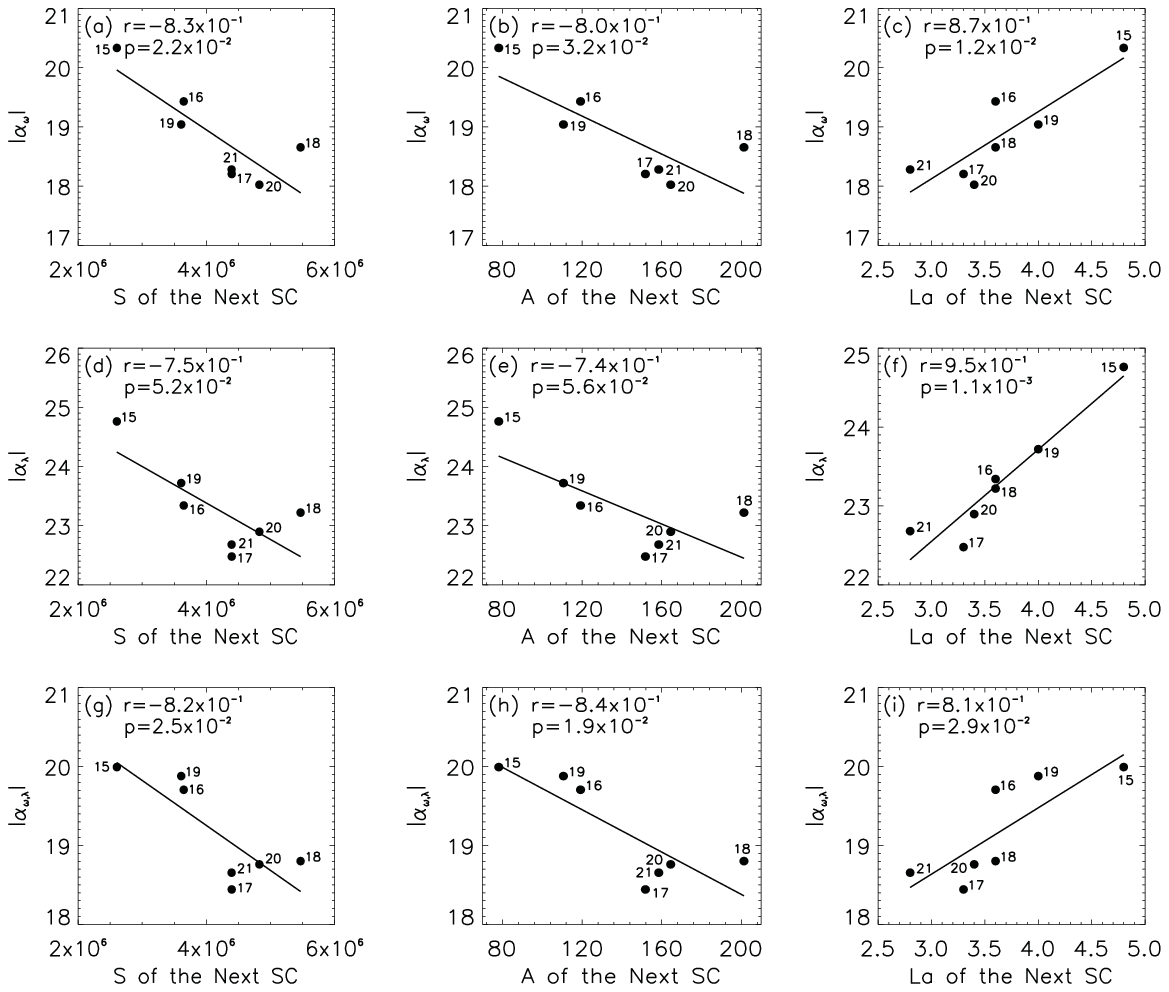}
      \caption{Panels (a), (b), and (c): the area-weighted absolute value of tilt angles during the declining phase vs. the strength, amplitude, and duration of the ascending phase of the next SC. Panels (d), (e), and (f): the latitude-weighted absolute value of tilt angles during the declining phase vs. the strength, amplitude, and duration of the ascending phase of the next SC. Panels (g), (h), and (i): the area- and latitude- weighted absolute value of tilt angles during the declining phase vs. the strength, amplitude, and duration of the ascending phase of the next SC. The solid lines show the respective linear fits. The correlation coefficients and probabilities can be found in each panel. These data are derived from the MW database.}
         \label{FigVibStab}
   \end{figure*}

The tilt-angle scatter during the declining phase exhibits significant anti-correlations with the strength and amplitude of the next SC, and a positive correlation with the duration of the ascending phase of the next SC (Table 6, Figure 12, and Figure 13).

Specifically, the standard deviation of tilt angles during the declining phase shows significant anti-correlations with the strength and amplitude of the next SC, $ r = -7.5\times10^{-1}$ at 94.6\% confidence level and $ r = -7.1\times10^{-1}$ at 92.6\% confidence level, respectively. It exhibits a positive correlation with the duration of the ascending phase of the next SC, $ r = 9.0\times10^{-1}$ at 99.37\% confidence level. Similarly, the root-mean-square tilt angle during the declining phase has significant anti-correlations with the strength and amplitude of the next SC, $ r = -7.9\times10^{-1}$ at 96.4\% confidence level and $ r = -7.4\times10^{-1}$ at 94.5\% confidence level, respectively, and a positive correlation with the duration of the ascending phase of the next SC, $ r = 8.7\times10^{-1}$ at 99.01\% confidence level. In addition, the mean absolute value of tilt angles during the declining phase exhibits significant anti-correlations with the strength and amplitude of the next SC, $ r = -7.5\times10^{-1}$ at 94.7\% confidence level and $ r = -6.9\times10^{-1}$ at 91.2\% confidence level, respectively. It also shows a positive correlation with the duration of the ascending phase of the next SC, $ r = 8.0\times10^{-1}$ at 97.0\% confidence level.

The area-weighted absolute value of tilt angles during the declining phase exhibits significant anti-correlations with the strength and amplitude of the next SC, $ r = -8.3\times10^{-1}$ at 97.8\% confidence level and $ r = -8.0\times10^{-1}$ at 96.8\% confidence level, respectively, and a positive correlation with the duration of the ascending phase of the next SC, $ r = 8.7\times10^{-1}$ at 98.8\% confidence level. Similarly, the latitude-weighted absolute value of tilt angles during the declining phase shows significant anti-correlations with the strength and amplitude of the next SC, $ r = -7.5\times10^{-1}$ at 94.8\% confidence level and $ r = -7.4\times10^{-1}$ at 94.4\% confidence level, respectively. It has a strong positive correlation with the duration of the ascending phase of the next SC, $ r = 9.5\times10^{-1}$ at 99.89\% confidence level. The area- and latitude- weighted absolute value of tilt angles during the declining phase exhibits significant anti-correlations with the strength and amplitude of the next SC, $ r = -8.2\times10^{-1}$ at 97.5\% confidence level and $ r = -8.4\times10^{-1}$ at 98.1\% confidence level, respectively, and a positive correlation with the duration of the ascending phase of the next SC, $ r = 8.1\times10^{-1}$ at 97.1\% confidence level.

The anti-correlation between the mean tilt angle during the declining phase and the duration of the ascending phase of the next SC is statistically insignificant. However, \citet{dasiespuig10} found that the anti-correlation between the mean tilt angle during the whole extent of a SC and the length of the next SC is statistically significant based on the MW data. Additionally, the positive correlation between the area-weighted tilt angle during the declining phase and the duration of the ascending phase of the next SC is statistically insignificant. The correlation between the area-weighted tilt angle during the whole extent of a SC and the length of the next SC is also statistically insignificant based on the MW data \citep{dasiespuig10}.

\subsubsection{The relationships of the tilt-angle scatter with the statistical properties of SGs}

\begin{figure*}
   \centering
   \includegraphics[bb=10 420 585 750, width=0.99\hsize]{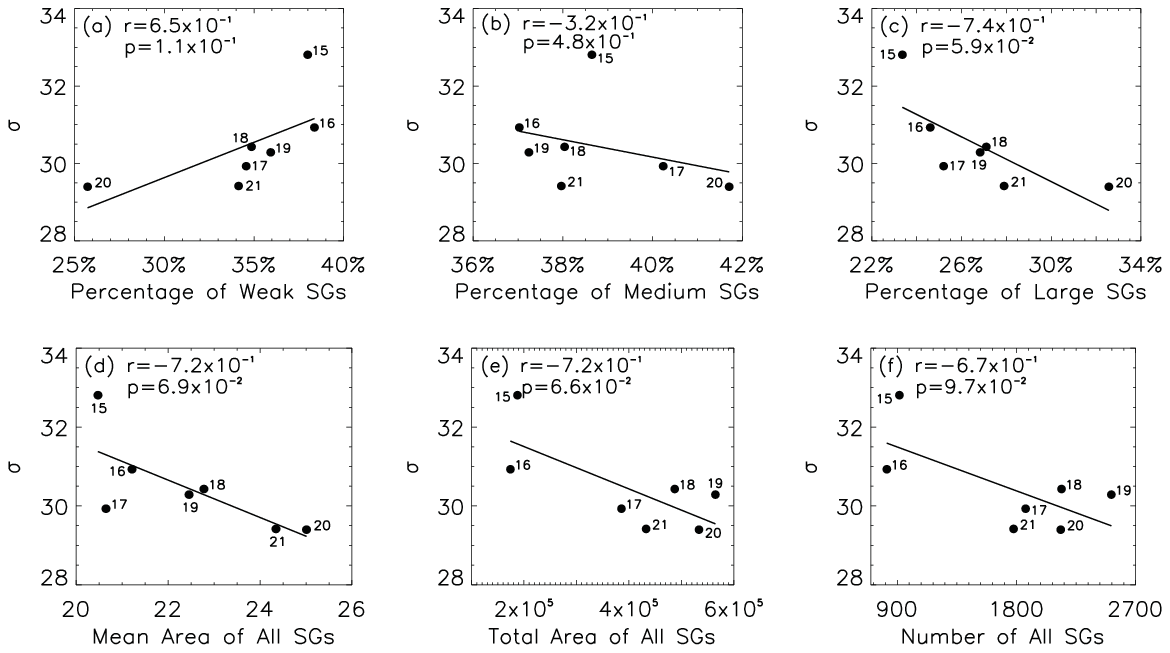}
      \caption{The standard deviation of tilt angles vs. the percentage of weak SGs (a), the percentage of medium SGs (b), the percentage of large SGs (c), the mean area of all SGs (d), the total area of all SGs (e), and the number of all SGs (f) during the declining phase. The solid lines show the respective linear fits. The correlation coefficients and probabilities can be found in each panel. These data are derived from the MW database.}
         \label{FigVibStab}
   \end{figure*}

\begin{table*}
 \centering
 \begin{minipage}{1.0\textwidth}
  \caption{The correlation coefficients of the standard deviation of tilt angles with the percentage of weak SGs, the percentage of medium SGs, the percentage of large SGs, the mean area of all SGs, the total area of all SGs, and the number of all SGs during the declining phase. These data are derived from the MW database.}
  \begin{tabular}{ccc}
  \hline
  \hline
Parameter & \multicolumn{2}{c}{$\sigma$} \\
          & r & p \\
\hline
Percentage of weak SGs                                     & $6.5\times10^{-1}$  & $1.1\times10^{-1}$ \\
Percentage of medium SGs                                   & $-3.2\times10^{-1}$ & $4.8\times10^{-1}$ \\
Percentage of large SGs                                    & $-7.4\times10^{-1}$ & $5.9\times10^{-2}$ \\
Mean area of all SGs                                       & $-7.2\times10^{-1}$ & $6.9\times10^{-2}$ \\
Total area of all SGs                                      & $-7.2\times10^{-1}$ & $6.6\times10^{-2}$ \\
Number of all SGs                                          & $-6.7\times10^{-1}$ & $9.7\times10^{-2}$ \\
\hline
\hline
\end{tabular}
\end{minipage}
\end{table*}

The standard deviation of tilt angles has significant anti-correlations with the percentage of large SGs, the mean area of all SGs, the total area of all SGs, and the number of all SGs, $ r = -7.4\times10^{-1}$ at 94.1\% confidence level, $ r = -7.2\times10^{-1}$ at 93.1\% confidence level, $ r = -7.2\times10^{-1}$ at 93.4\% confidence level, and $ r = -6.7\times10^{-1}$ at 90.3\% confidence level, respectively (Table 7 and Figure 14).

\subsubsection{The relationships of the tilt angle during the declining phase with the parameters of the same SC.}

\begin{table*}
 \centering
 \begin{minipage}{1.0\textwidth}
  \caption{The correlation coefficients of the 10 quantities based on the tilt angles during the declining phase with the strength (S), amplitude (A), and duration of the ascending phase (La) of the same SC. These data are derived from the MW database.}
  \begin{tabular}{ccccccc}
  \hline
  \hline
Parameter & \multicolumn{2}{c}{S} & \multicolumn{2}{c}{A} & \multicolumn{2}{c}{La}\\
          & r & p & r & p & r & p\\
\hline
$\langle\alpha\rangle$      & $3.5\times10^{-1}$  & $4.4\times10^{-1}$ & $3.1\times10^{-1}$  & $5.0\times10^{-1}$ & $-5.8\times10^{-2}$ & $9.0\times10^{-1}$ \\
$\alpha_\omega$             & $-4.8\times10^{-1}$ & $2.8\times10^{-1}$ & $-3.6\times10^{-1}$ & $4.3\times10^{-1}$ & $3.2\times10^{-1}$  & $4.9\times10^{-1}$ \\
$\alpha_\lambda$            & $2.7\times10^{-1}$  & $5.5\times10^{-1}$ & $2.4\times10^{-1}$  & $6.1\times10^{-1}$ & $7.5\times10^{-2}$  & $8.7\times10^{-1}$ \\
$\alpha_{\omega,\lambda}$   & $-3.8\times10^{-1}$ & $4.0\times10^{-1}$ & $-3.5\times10^{-1}$ & $4.4\times10^{-1}$ & $4.1\times10^{-1}$  & $3.6\times10^{-1}$ \\
\hline
$\sigma$                    & $-6.2\times10^{-1}$ & $1.4\times10^{-1}$ & $-3.6\times10^{-1}$ & $4.3\times10^{-1}$ & $3.6\times10^{-1}$  & $4.3\times10^{-1}$ \\
RMS                         & $-6.0\times10^{-1}$ & $1.5\times10^{-1}$ & $-3.3\times10^{-1}$ & $4.7\times10^{-1}$ & $3.6\times10^{-1}$  & $4.3\times10^{-1}$ \\
$\langle|\alpha|\rangle$    & $-5.0\times10^{-1}$ & $2.5\times10^{-1}$ & $-2.3\times10^{-1}$ & $6.2\times10^{-1}$ & $2.4\times10^{-1}$  & $6.0\times10^{-1}$ \\
$|\alpha_\omega|$           & $-5.2\times10^{-1}$ & $2.3\times10^{-1}$ & $-2.6\times10^{-1}$ & $5.7\times10^{-1}$ & $4.5\times10^{-1}$  & $3.1\times10^{-1}$ \\
$|\alpha_\lambda|$          & $-3.6\times10^{-1}$ & $4.3\times10^{-1}$ & $-7.6\times10^{-2}$ & $8.7\times10^{-1}$ & $2.6\times10^{-1}$  & $5.7\times10^{-1}$ \\
$|\alpha_{\omega,\lambda}|$ & $-2.6\times10^{-1}$ & $5.7\times10^{-1}$ & $-4.7\times10^{-2}$ & $9.2\times10^{-1}$ & $5.0\times10^{-1}$  & $2.5\times10^{-1}$ \\
\hline
\hline
\end{tabular}
\end{minipage}
\end{table*}

The correlations of 10 quantities measured during the declining phase with the strength, amplitude, and duration of the ascending phase of the same SC are statistically insignificant: the mean tilt angle, the area-weighted tilt angle, the latitude-weighted tilt angle, the area- and latitude- weighted tilt angle, the standard deviation of tilt angles, the root-mean-square tilt angle, the mean absolute value of tilt angles, the area-weighted absolute value of tilt angles, the latitude-weighted absolute value of tilt angles, and the area- and latitude- weighted absolute value of tilt angles. The anti-correlations of the mean tilt angle and the area-weighted tilt angle during the whole extent of a SC with the strength, amplitude, and length of the same SC also are statistically insignificant based on the MW data \citep{dasiespuig10}.

\section{Conclusions and discussion} \label{sec:highlight}

Based on the data from the KK and MW observatories, we investigate the relationships of the tilt angle, including the mean tilt angle and the tilt-angle scatter, during the declining phase with the SC parameters. The main findings can be summarized into three points:

\begin{itemize}

\item During the declining phase, the correlation between the mean tilt angle and the tilt-angle scatter is statistically insignificant.

\item Six quantities measured during the declining phase show significant anti-correlations with the strength and amplitude of the next SC, and positive correlations with the duration of the ascending phase of the next SC: the standard deviation of tilt angles, the root-mean-square tilt angle, the mean absolute value of tilt angles, the area-weighted absolute value of tilt angles, the latitude-weighted absolute value of tilt angles, and the area- and latitude- weighted absolute value of tilt angles. An exception is the correlation of the area- and latitude- weighted absolute value of tilt angles during the declining phase with the duration of the ascending phase of the next SC for the KK data.

\item The correlations of the mean tilt angle, the area-weighted tilt angle, the latitude-weighted tilt angle, and the area- and latitude- weighted tilt angle during the declining phase with the strength, amplitude, and duration of the ascending phase of the next SC are statistically insignificant. The exception is the correlation of the area- and latitude- weighted tilt angle during the declining phase with the amplitude of the next SC without SC 16 (an outlier) for the KK data.

\end{itemize}

These findings suggest that the tilt-angle scatter during the declining phase is an essential ingredient of the solar dynamo. The modulation of parameters of the next SC by the tilt-angle scatter during the declining phase plays a vital role in regulating SC variability, consistent with the results of numerical simulations: the tilt-angle scatter significantly impacts the strength of the next SC \citep{baumann04,jiang14,bhowmik18}.

The statistically insignificant correlation between the mean tilt angle and the tilt-angle scatter may be the reason why they show different behaviors, as previously noted by \citet{wang89} and \citet{gao23}.

\citet{bogdan88} reported that the sunspot umbral areas are distributed lognormally and the distributions remain constant for various individual SCs. However, \citet{tang84} considered that the mean area of SGs is an indicator of SC variation. Our findings show that the mean area of SGs during the declining phases of SCs 14-21 varies from one SC to the next, supporting the conclusions of \citet{tang84} and \citet{li05a}. Additionally, the total mean of SGs is also an indicator of SC variation \citep{li05a,balmaceda09, hathaway15,norton23}.

During the declining phase, the standard deviation of tilt angles (tilt-angle scatter) shows a significant anti-correlation with the mean area of SGs for the MW and KK (excluding SC 15,  an outlier) data. It also has significant anti-correlations with the total area of SGs and the percentage of large SGs for the MW and KK (with and without SC 15) data, supporting the viewpoint that the tilt-angle scatter decreases as SG area increas \citep{wang89,jiang14}. To some extent, SG area serves as a proxy for total magnetic flux \citep{zharkov06, dikpati06}. The tilt-angle scatter during the declining phase has significant anti-correlations with the strength and amplitude of the next SC. These results suggest that the weaker the solar activity during the declining phase, the weaker the solar activity during the subsequent SC, and vice versa.

The tilt-angle scatter during the declining phase exhibits significant anti-correlations with the strength and amplitude of the next SC, while simultaneously showing a significant positive correlation with the duration of the ascending phase of the next SC. Hence, the strength and amplitude of a given SC are anti-correlated with the duration of its ascending phase (Waldmeier effect, \citet{waldmeier35}), which forms the basis of some methods used to predict the amplitude of SC \citep{wang02,li05b}.

The anti-correlations of the area-weighted absolute value of tilt angles during the declining phase with the strength and amplitude of the next SC indicates that SGs with larger areas generally contribute more to SC variability. This is likely because SG area, to some extent, serves as a proxy for total magnetic flux \citep{zharkov06, dikpati06}.

Previous studies suggest that low-latitude SGs can significantly impact the evolution of polar field \citep{cameron13,jiang15,nagy17}. As is known, SGs appear at lower latitudes during the declining phase. Therefore, compared with SGs during the rising and maximum phases, which appear at higher latitudes, SGs during the declining phase should have a greater impact on the evolution of polar fields. Low-latitude magnetic flux is transported to the pole taking about 2 years \citep{jiang14}. Thus, the tilt angles of SGs during the declining phase (after the polarity reversal) are more crucial for the build-up of the opposite-polarity polar field, which is the source of the toroidal field responsible for solar activity during the next SC. These may partly explain why we find that the tilt-angle scatter during the declining phase shows significant anti-correlations with the strength and amplitude of the next SC.

\begin{figure*}
   \centering
   \includegraphics[bb=10 585 580 745, width=0.99\hsize]{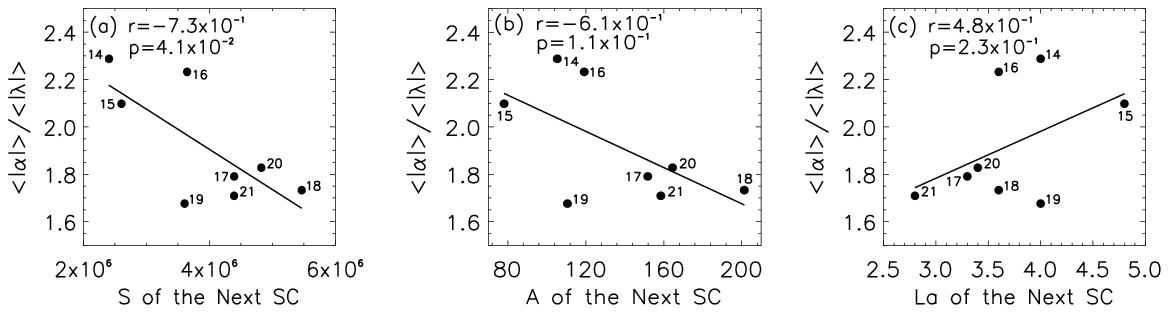}
      \caption{$\langle|\alpha|\rangle/\langle|\lambda|\rangle$ during the declining phase vs. the strength (a), amplitude (b), and duration of the ascending phase (c) of the next SC. The solid lines show the respective linear fits. The correlation coefficients and probabilities can be found in each panel. These data are derived from the KK database.}
         \label{FigVibStab}
   \end{figure*}

However, we also find that the latitude-weighted absolute value of tilt angles during the declining phase is significantly anti-correlated with the strength and amplitude of the next SC, indicating that, for forecasting solar activity during the next SC, the higher the latitude of SGs during the declining phase, the greater the contribution. This is contradictory to the results of previous studies --- the low-latitude SGs can have a major impact on the evolution of polar fields \citep{cameron13,jiang15,nagy17}.

Therefore, we investigate the relationships of $\langle|\alpha|\rangle/\langle|\lambda|\rangle$ during the declining phase with the strength, amplitude, and duration of the ascending phase of the next SC based on the KK data (Figure 15). The $\langle|\alpha|\rangle/\langle|\lambda|\rangle$ during the declining phase shows a significant anti-correlation with the strength of the next SC, $ r = -7.3\times10^{-1}$ at 95.9\% confidence level. The correlations of the $\langle|\alpha|\rangle/\langle|\lambda|\rangle$ during the declining phase with the amplitude and duration of the ascending phase of the next SC are statistically insignificant. Further efforts are underway to understand the role of SG latitude in the mechanism of modulation of the parameters of the next SC.

It must be noted that the tilt angle of SG is significantly different in the Northern and Southern hemispheres \citep{mcclintock13,li12,javaraiah23}. Besides, a north-south asymmetry also exists in the SC properties \citep{norton10,mcclintock13,mcintosh13,javaraiah19,ravindra21}. To gain a more comprehensive understanding of the operational mechanism of the solar dynamo, we will investigate the relationships between the tilt angle and the SC parameters in the Northern and Southern hemispheres, respectively.

\acknowledgments
The authors thank the reviewer very much for the careful reading and the constructive comments that improved the original version of the manuscript. This work is supported by the National Natural Science Foundation of China (Grant Nos. 12373061), the Basic Research Foundation of Yunnan Province, China (Grant Nos. 202101AT070019, 202201AS070042), Yunnan Key Laboratory of Solar Physics and Space Science (Grant Nos. 202205AG070009), the Basic Research Priorities Program of Yunnan (202401CF070062), and the Chinese Academy of Sciences.

\end{document}